\documentclass[twocolumn,english,nobibnotes,nofootinbib]{revtex4-1}
\usepackage[T1]{fontenc}
\usepackage[a4paper]{geometry}
\usepackage{float}
\usepackage{textcomp}
\usepackage{amsmath}
\usepackage{amssymb}
\usepackage{graphicx}
\usepackage{esint}
\usepackage{natbib}
\usepackage{color}

\makeatletter
\@ifundefined{textcolor}{}
{%
 \definecolor{BLACK}{gray}{0}
 \definecolor{WHITE}{gray}{1}
 \definecolor{RED}{rgb}{1,0,0}
 \definecolor{GREEN}{rgb}{0,1,0}
 \definecolor{BLUE}{rgb}{0,0,1}
 \definecolor{CYAN}{cmyk}{1,0,0,0}
 \definecolor{MAGENTA}{cmyk}{0,1,0,0}
 \definecolor{YELLOW}{cmyk}{0,0,1,0}
}

\begin{document}

\title{Imaging spin-resolved cyclotron trajectories in the InSb two-dimensional
electron gas }
\author{A. Mre\'{n}ca-Kolasi\'{n}ska}
\affiliation{AGH University of Science and Technology, Faculty of Physics and
Applied Computer Science,\\
 al. Mickiewicza 30, 30-059 Krak\'ow, Poland}

\author{K. Kolasi\'{n}ski }
\affiliation{AGH University of Science and Technology, Faculty of Physics and
Applied Computer Science,\\
 al. Mickiewicza 30, 30-059 Krak\'ow, Poland}
 
\author{B. Szafran}
\affiliation{AGH University of Science and Technology, Faculty of Physics and
Applied Computer Science,\\
 al. Mickiewicza 30, 30-059 Krak\'ow, Poland}
\begin{abstract}
We consider spin-resolved cyclotron trajectories in 
a magnetic focusing device with quantum point
source and drain contacts defined within the InSb two-dimensional electron gas and their mapping by the scanning gate microscopy.  
Besides the perpendicular component of the external magnetic field which
bends the electron trajectories we consider an in-plane component which
introduces the spin-dependence of the cyclotron radius. 
We demonstrate that the focusing conductance peaks become spin-split by
the in-plane magnetic field component of the order of a few tesla and that the
spin-resolved trajectories can be traced separately with the conductance mapping. 
\end{abstract}
\maketitle

\section{Introduction}


The separation and control of the electron spins in the two-dimensional electron gas (2DEG) has been a subject of intense investigation in the field of spintronics \cite{Wolf2001}. 
In the external magnetic field the focusing 
of the cyclotron trajectories can be detected in a set-up with  quantum point contact (QPC) source and drain terminals \cite{Sharvin1964, Tsoi1974, Houten1989,Hanson2003,Aidala2007,Dedigama2006, Lo2017, Yan2017,Rokhinson2004, Chesi2011, Rokhinson2006}.
In this work we consider the spin-dependent trajectories that could be resolved in the magnetic focusing experiment
\cite{Sharvin1964, Tsoi1974, Houten1989,Hanson2003,Aidala2007,
Dedigama2006, Lo2017, Yan2017,Rokhinson2004, Chesi2011, Rokhinson2006}
by the scanning gate microscopy \cite{Sellier2011}. 
The focusing of electron  trajectories for carriers injected across the QPC with spins 
separated by spin-orbit interaction (SOI)
was considered theoretically \cite{Usaj2004, Zulicke2007, Reynoso2007, Schliemann2008, Reynoso2008b, Kormanyos2010, Bladwell2015} and 
studied experimentally \cite{Rokhinson2004, Dedigama2006, Chesi2011, Lo2017, Yan2017, Yan2018}.
The spin separation by the strong spin-orbit interaction is achieved by splitting the magnetic focusing peaks with the orthogonal spin polarization for electrons that pass across the quantum point contacts.
The spin-orbit coupling alone in the absence of the external magnetic field has also been proposed
for the spin-separation in InGaAs QPCs \cite{Kohda2012} and 
in U- \cite{Zeng2012} or Y-shaped \cite{Gupta2014} junctions of topological insulators.
 However, for strong spin-orbit coupling the electron spin precesses in the 
effective momentum-dependent spin-orbit magnetic field \cite{Meier2007,Reynoso2008} that is oriented within the plane of confinement of the carrier gas.
In this work we indicate a possibility of imaging the spin-resolved electron trajectories
for which the electron spin is fixed and the spin-precession in the spin-orbit field is frozen
by strong Zeeman effect due to an in-plane magnetic field. 
  For that purpose instead of the spin-orbit coupling \cite{Usaj2004, Schliemann2008, Kormanyos2010,Rokhinson2004, Dedigama2006, Chesi2011, Lo2017, Yan2017}  we use an in-plane magnetic field \cite{Watson2003,Li2012,Yan2018a} component
that introduces the spin-dependence of the cyclotron trajectories by the Zeeman splitting. 
We demonstrate that for the indium antimonide -- a  large Land\'e factor material --the spin-dependent electron trajectories can be clearly resolved by the scanning gate microscopy technique. 

In the focusing experiments with the 2DEG the electrons are injected and gathered by QPCs \cite{Wharam1988, Wees1988, Wees1991}. The constrictions formed in 2DEG by electrostatic gates depleting the electron gas lead to the formation of transverse quantized modes. By applying sufficiently high potential on the gates only one or few modes can adiabatically pass through the QPC. The quantized plateaus of conductance of such constrictions have been recently reported in InSb \cite{Qu2016}.

The scanning gate microscopy (SGM) is an experimental technique in which a charged tip of atomic force microscope is raster-scanned over a sample while measuring the conductance \cite{Sellier2011}. The tip acts as a movable gate that can locally deplete the 2DEG, with a possible effect on the conductance. 
The SGM technique has been used in 2DEG confined in III-V nanostructures for example to image the branching of the current trajectories in systems with QPC and the interference of electrons backscattered between the tip and the QPC \cite{LeRoy2005, Jura2009, Paradiso2010, Brun2014}, the scarred wave functions in quantum billiards \cite{Crook2003, Burke2010}, 
and electron cyclotron trajectories \cite{Aidala2007, Crook2010}. It has been used for imaging the cyclotron motion also in two-dimensional materials like graphene \cite{Morikawa2015, Bhandari2016}. 


%

\section{Model and theory}
 
\begin{figure}[tb!]
 \includegraphics[width=\columnwidth]{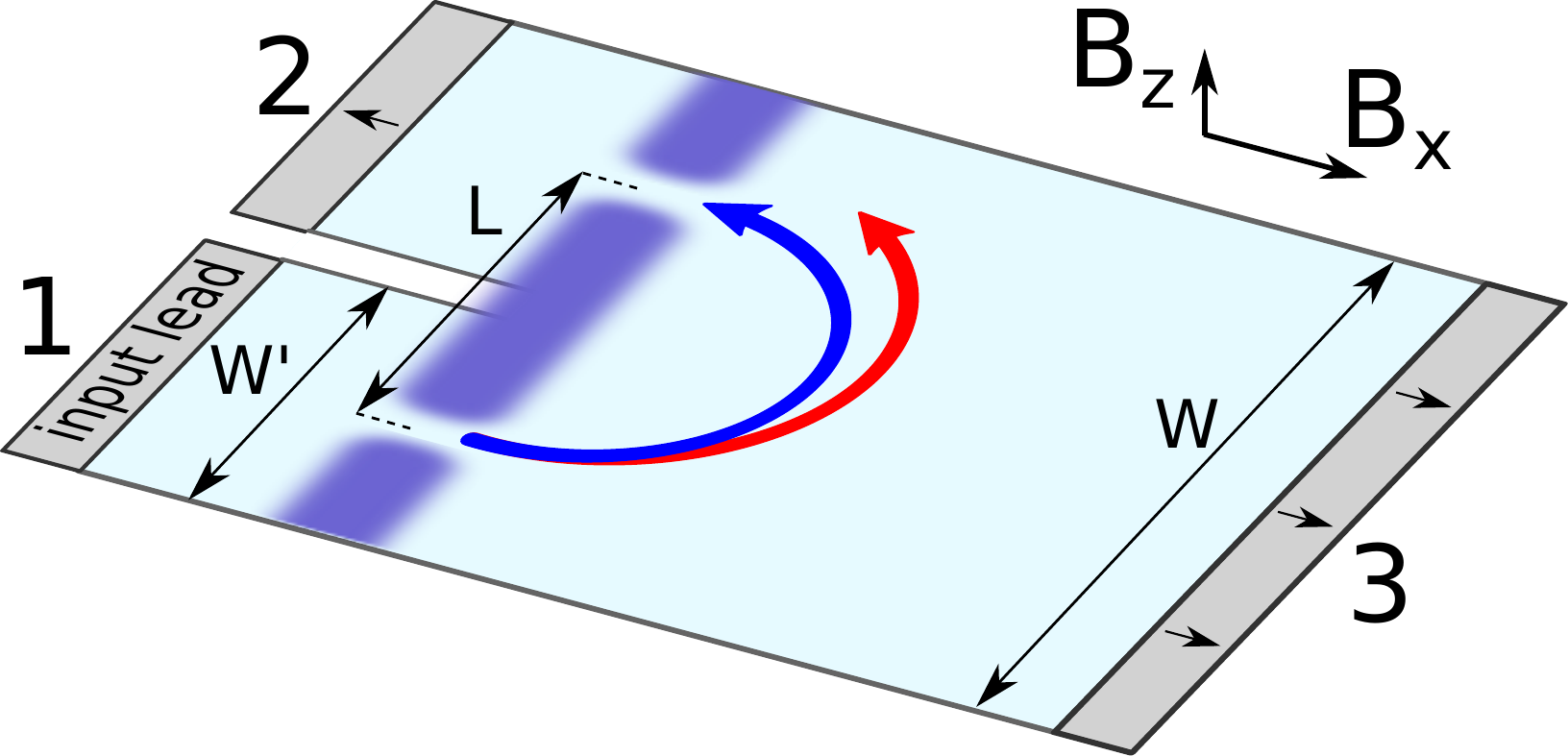}
  \caption{The scheme of the focusing system. 
The dark blue shaded area is the gate-induced potential defining the two QPCs, separated by the distance $L$. The spin up (spin down) is parallel (antiparallel) to the total magnetic field. Due to the in-plane magnetic field (and hence Zeeman splitting) the spin-up and spin-down electrons have different momenta and get spatially separated due to difference in the cyclotron radii. The red and blue arrows correspond to spin-up and spin-down electron trajectories, respectively. The gray rectangles indicate the open boundary conditions. The terminals are numbered by integers from 1 to 3. Terminal 1 (2) is the source (drain) of the currents. Terminal 3 plays a role of an open boundary. 
  } \label{system}
\end{figure}

We consider quantum transport at the Fermi level in 2DEG confined within an InSb quantum well. 
The model system depicted in Fig.~\ref{system} contains two QPCs on the left-hand side, and is open on the right-hand side. The electrostatically defined two quantum point contacts are separated by a distance $L$. The terminals are numbered as indicated in Fig.~\ref{system}. The electrons entering from the lead 1 are injected trough the first (lower) QPC into the system in a narrow beam that is steered by the transverse magnetic field. Whenever the cyclotron diameter (or its integer multiple) fits the separation $L$, electrons can enter the second QPC which serves as a collector. 
Electrons that do not get to the collector exit the system through the 
lead 3, which is used as open boundary conditions. Hard wall boundary conditions are introduced on the perpendicular edges of the computational box. The size of the computational box (width $W=2400$ nm and length 1800 nm) 
is large enough to make the effects of the scattering by the hard wall boundaries negligible for the drain (lead 2) currents.

For the transport modeling, we
assume that the vertical confinement in the InSb quantum well
is strong enough to justify the two-dimensional approximation for
the electron motion. The 2D effective mass Hamiltonian reads
\begin{eqnarray}
H=& \left[\frac{\hbar^2}{2m_{eff}}\mathbf{k}^2 + eV(\mathbf{r}) \right]\mathbf{1} +\frac{1}{2}\mu_B \boldsymbol{B}^T \boldsymbol{g}^* \boldsymbol{\sigma} +H_{SO}, 
\label{eq:dh}
\end{eqnarray}
where $\mathbf{k}=-i\boldsymbol{\nabla}-e\mathbf{A}$, with $\mathbf{A}$ being the vector potential, $\mathbf{B}=(B_x,B_y,B_z)$, $\boldsymbol{\sigma}$ is the vector of Pauli matrices, $\mu_B$ is the Bohr  magneton, $\mathbf{g}^*$ is the diagonal Land\'e tensor, and $m_{eff}$ is the electron effective mass in InSb.

\begin{figure}[tb!]
 \includegraphics[width=0.6\columnwidth]{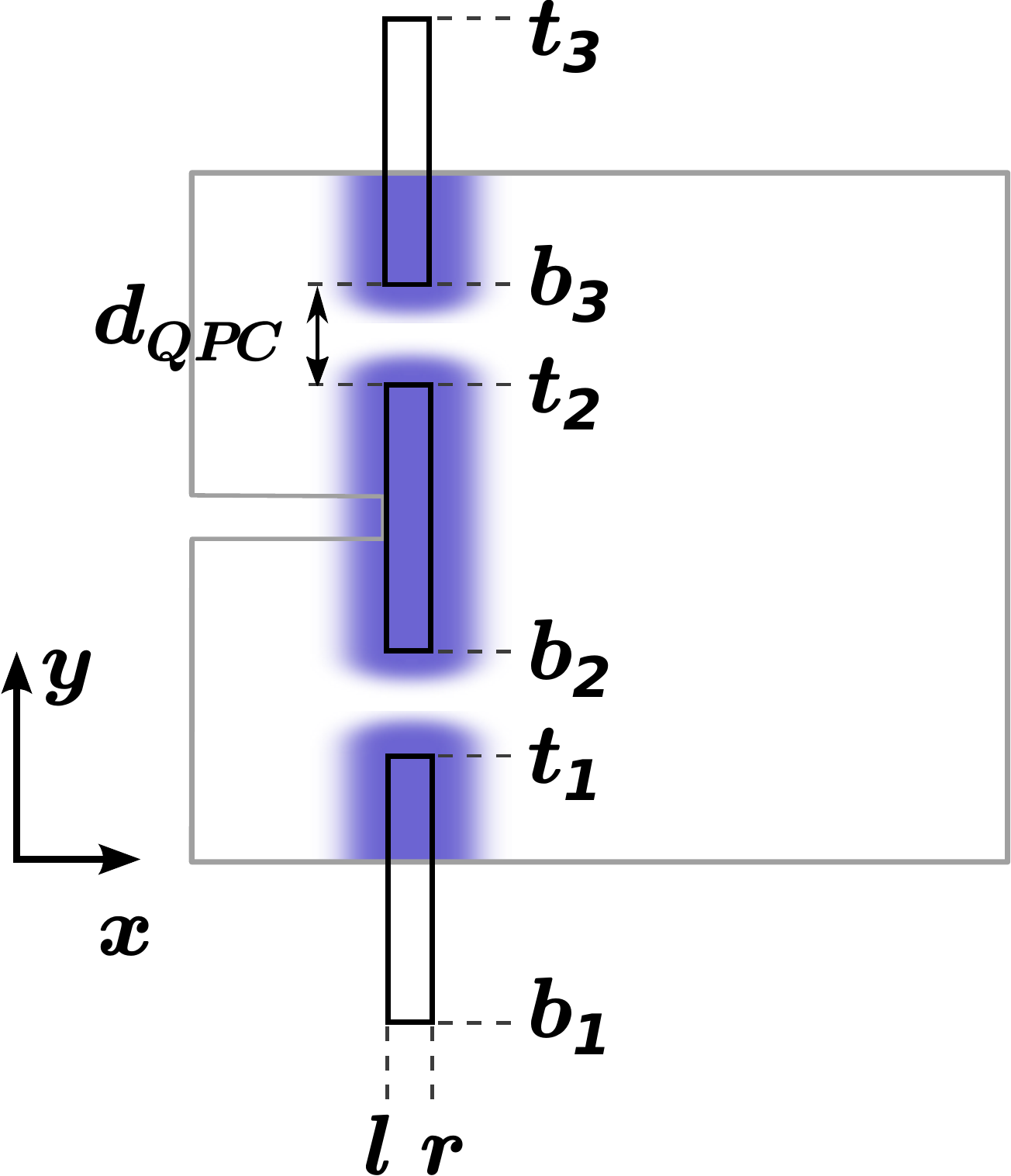}
  \caption{The scheme of the gates inducing the potential of the two QPCs. 
The figure is not to scale. The values of the geometrical parameters are:  $l=300$ nm, $r=500$ nm, $b_1=-600$ nm, $t_1=547$ nm, $t_2=652$ nm, $b_2=1747$ nm, $b_3=1852$ nm, and $t_3=3000$ nm.}.
   \label{gates}
\end{figure}

The external potential as seen by the Fermi level electrons is a superposition of the QPC and the potential induced by the charged SGM tip 
\begin{equation}
V(\mathbf{r}) = V_{QPC}(\mathbf{r})+V_{tip}(\mathbf{r})  , 
\label{eq:Vext}
\end{equation}
where we model the QPC using the analytical formulas developed in \cite{Davies1995} with electrostatic potential of a finite rectangular gate given by 
\begin{eqnarray}
\begin{aligned}
 V_{r}(\mathbf{r};l,r,b,t)=\\
 V_g &\left[g( x-l,y-b  ) + g( x-l,t-y ) \right. \\
 +& \left. g( r-x,y-b ) +g( r-x,t-y)  \right],
 \end{aligned}
\end{eqnarray}
where $g(u,v) = \tfrac{1}{2\pi} \arctan\left( \tfrac{uv}{u^2+v^2+d^2} \right)$ with $d=50$ nm, and $V_g$ is the potential applied to the gates. The QPC potential is a superposition of potentials of three such gates
\begin{equation}
 V_{QPC} = V_{r}(\mathbf{r};l,r,b_1,t_1) + V_{r}(\mathbf{r};l,r,b_2,t_2) + V_{r}(\mathbf{r};l,r,b_3,t_3).
 \label{eq:qpc_gates}
\end{equation}
 The gates and their labeling used in Eq.~(\ref{eq:qpc_gates}) are schematically shown in Fig.~\ref{gates}. 
The splitting of the gates is $d_{QPC}=105$ nm defining the QPC width. The QPCs are separated by $L=1200$ nm. 

For modeling the tip potential we use a Gaussian profile
\begin{equation}
 V_{tip}(\mathbf{r})= V_t \exp \left[ -\frac{(x-x_{tip})^2+(y-y_{tip})^2}{d_{tip}^2} \right],
\end{equation}
with $V_{t}$ being the maximum tip potential, $d_{tip}$ its width, and $x_{tip}$, $y_{tip}$ the coordinates of the tip.

The spin-orbit interactions in InSb are strong, so we include them in the calculations.
The two last terms in (\ref{eq:dh}) account for the SOI with $H_{SO} =H_{R} + H_{D}$, where 
\begin{equation}
H_{R} = \alpha( - k_x\sigma_y + k_y\sigma_x )  
\label{eq:HR}
\end{equation}
describes the Rashba interaction, and 
\begin{equation}
H_{D} = \beta( k_x\sigma_x - k_y\sigma_y )  
\label{eq:HD}
\end{equation}
the Dresselhaus interaction. 
For the Hamiltonian (\ref{eq:dh}) we use the parameters for InSb quantum well, $\alpha=-0.051$ eV\AA, $\beta=0.032$ eV\AA, $g^*_{zz}=-51$ \cite{Gilbertson2008}, $g^*_{xx}=\tfrac{1}{2}g^*_{zz}$ \cite{Qu2016}, $m_{eff}=0.018m_0$ \cite{Qu2016}.

We perform the transport calculations in the finite difference formalism. For evaluation of the transmission probability, we use the wave function matching (WFM) technique \cite{Kolacha}. The transmission probability from the input lead to mode $m$ with spin $\sigma$ in the output lead is
\begin{equation}
T^m_\sigma = \sum_{ n,\sigma'} |t^{mn}_{\sigma\sigma'}|^2,
\label{eq:transprob}
\end{equation}
where $t^{mn}_{\sigma\sigma'}$ is the probability amplitude for the transmission from the mode $n$ with spin $\sigma'$ in the input lead to mode $m$ with spin $\sigma$ in the output lead. 
We evaluate the conductance as $G={G_0}\sum_{m, \sigma} T^{m}_\sigma$, with $G_0={e^2}/{h}$.

The considered system presented in Fig.~\ref{system} has the width $W=2400$ nm, and the narrow leads numbered 1 and 2 have equal width $W'=1146$ nm. The spacing between the centers of the QPCs is $L=1200$ nm. We take the gate potential $V_g=62$ meV, for which at  $E_F=26$ meV
in the absence of the external magnetic field the QPC conductance is close to $2\tfrac{e^2}{h}$. For the SGM we use the tip parameters $V_t=260$ meV, and $d_{tip}=60$ nm.

\section{Results}

\subsection{No in-plane magnetic field}

\begin{figure}[tb!]
 \includegraphics[width=\columnwidth]{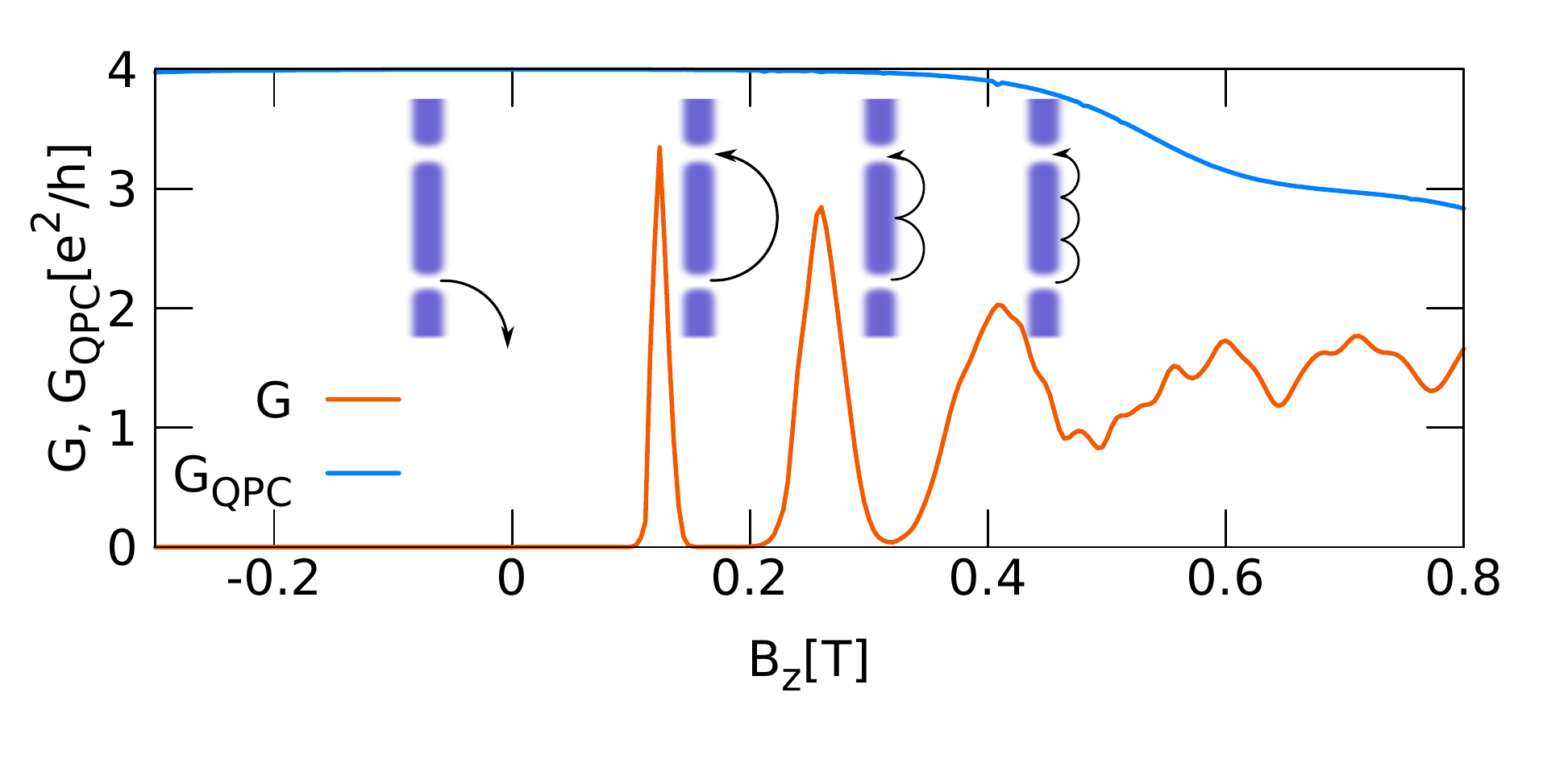}
  \caption{The conductance from the left bottom to the left top lead $G$ as a function of magnetic field and the lower QPC conductance $G_{QPC}$. The inset shows semi-classical trajectories of the electrons for $B_z<0$, and at the three focusing peaks $B_{z}^{(i)}$ with $i$=1,2,3.
  } \label{fig:onlyBz}
\end{figure}

Let us first consider the transport in the system with the out-of-plane magnetic field only (i.e. $B_x=0$, $B_y=0$, $B_z \ne 0$). 
In Fig.~\ref{fig:onlyBz} we present the conductance $G=G_{21}$ from the lead 1 to lead 2 as a function of the applied transverse magnetic field, and the summed conductance from the lead 1 to the leads 2 and 3, which is essentially the conductance of the lower QPC $G_{QPC}=G_{21}+G_{31}$.
For $B_z<0$ no focusing peaks occur because the electrons are deflected in the opposite direction than the collector, propagate along the bottom edge of the system and finally exit through the right lead. For $B_z>0$ conductance peaks almost equidistant in magnetic field appear. The first three maxima occur at $B_z^{(1)}=0.124$ T, $B_z^{(2)}=0.26$ T, $B_z^{(3)}=0.408$ T. Neglecting the SOI terms and the Zeeman term in (\ref{eq:dh}), one obtains $|k_F|=\sqrt{2m_{eff}E_F}=0.2148 \tfrac{1}{\mathrm{nm}}$. For the cyclotron diameter equal to 
\begin{equation}
D_c=\frac{2\hbar |k_F|}{|e| B_z}, 
\label{eq:Dc}
\end{equation}
one obtains for the first three peaks $D_c^{(1)}=1176$ nm, $D_c^{(2)}=561$ nm, $D_c^{(3)}=358$ nm, respectively. This is close to the distance between the centers of the QPCs, $L=$1200 nm, its half , $L/2=$600 nm, and one third, $L/3=$400 nm, respectively. 
Despite the high spin-orbit interaction in the InSb quantum well, no spin splitting occurs. Let us denote the Fermi wave number of the subband of spin $\sigma$ by $k_F^{\sigma}$. For the adapted values of the SO parameters, the difference in momenta for both spins is small [see Fig.~\ref{fig:disp2dBx}(a)]. For example for $k_{F,y}^{\uparrow},k_{F,y}^{\downarrow}=0$ and $E_F=26$ meV, the $x$ components extracted from the dispersion relation in Fig.~\ref{fig:disp2dBx}(a) are $|k_{F,x}^\uparrow|=0.11445$ nm$^{-1}$, and $|k_{F,x}^\downarrow|=0.11142$ nm$^{-1}$, that for $D_c=1200$ nm yield transverse magnetic field $B_z^{(1)}=0.125$ T and $B_z^{(1)}=0.122$ T, respectively. 
That is clearly too small difference to obtain a visible double peak. 


\subsection{Enhancement of the Zeeman splitting with in-plane magnetic field}

\begin{figure}[tb!]
 \includegraphics[width=0.49\columnwidth]{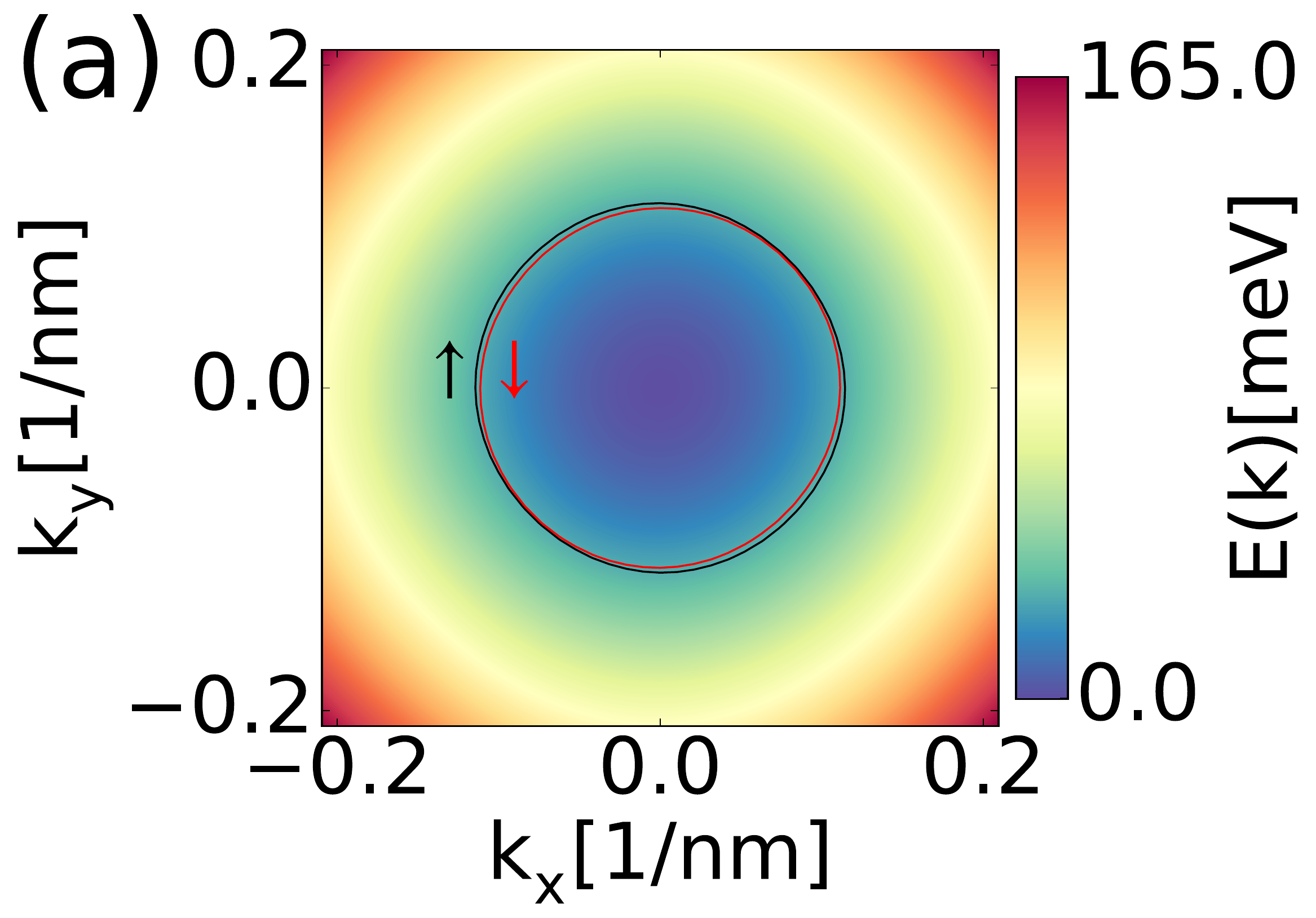}
 \includegraphics[width=0.49\columnwidth]{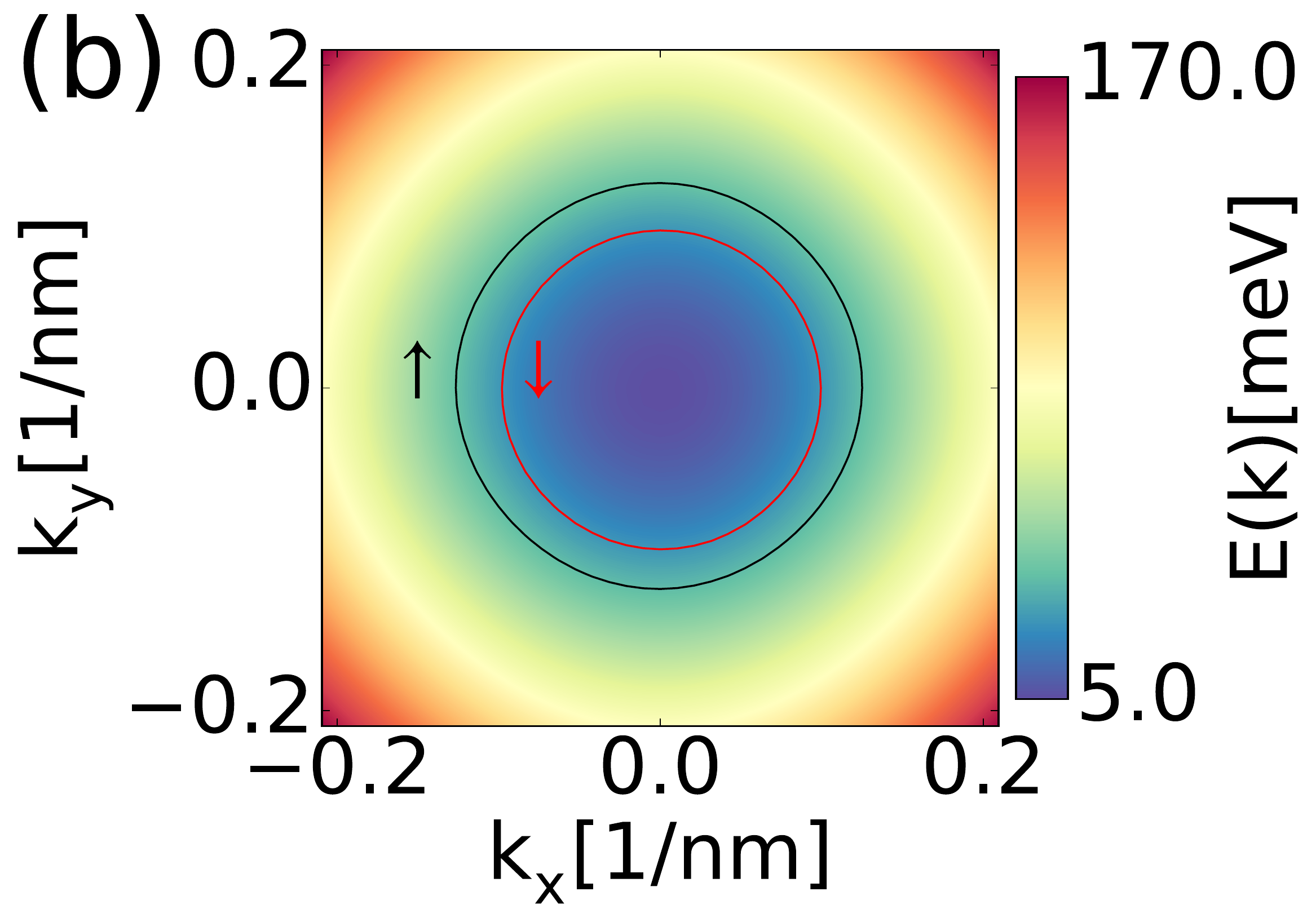}
  \caption{Dispersion relation of the 2DEG with (a) $B_x=0$ and (b) $B_x=8$ T. The color map shows the dispersion relation of the spin-down band, and the contours show the isoenergetic lines for $E_f=26$ meV for spin up (black line) and spin down (red line) electrons.
  } \label{fig:disp2dBx}
\end{figure}

In the next step we apply an additional in-plane magnetic field. This leads to an increase of the Zeeman energy splitting for both spins leading to the increase of the momenta difference between both spin subbands. Fig.~\ref{fig:disp2dBx} shows the momenta for both spins for $B_x=0$ and 8 T. Without in-plane magnetic field, the spin subbands are nearly degenerate. With $B_x$ of the order of a few tesla the difference in the momenta becomes significant. This induces a change of the cyclotron radii of the electrons with opposite spins. 

The spins are oriented along the total magnetic field, $\mathbf{B}+\mathbf{B}_{SO}$, where $\mathbf{B}_{SO}$ is the effective SO field. For $B_x$ of the order of a few tesla the out-of-plane magnetic field component and the SO effective field are small compared to the in-plane component. The spin is oriented nearly along the $x$ or $-x$ direction. We refer to these states as spin-up and spin-down.

\begin{figure}[tb!]
 \includegraphics[width=\columnwidth]{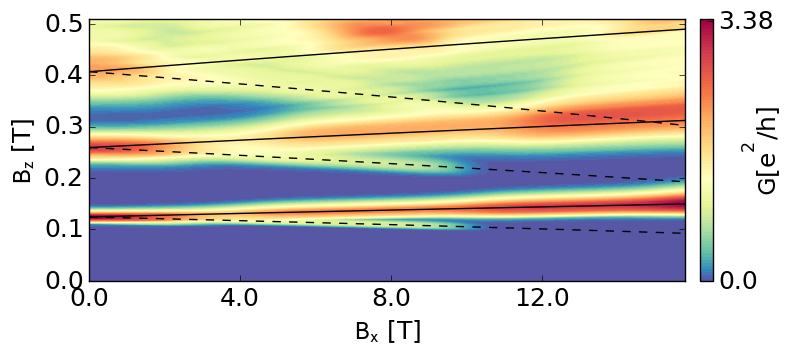}
  \caption{Transmission as a function of $B_x$ and $B_z$. The solid (dashed) lines are the analytically calculated positions of transmission peaks maxima for spin up (down) electrons.
  } \label{fig:transpBx}
\end{figure}

Fig.~\ref{fig:transpBx} shows the conductance $G$ from the lead 1 to the lead 2 as a function of the in-plane (here $B_x$) and the transverse magnetic fields. For sufficiently high in-plane magnetic field the peaks split, with the splitting growing with increasing $B_x$. The lines plotted along the $n$-th pair of split peaks are calculated from the condition $B_{z,\sigma}^{(n)}\left(B_x\right)=\tfrac{2\hbar |k_F^\sigma|}{|e| D_c^{(n)}}$, with  $|k_F^\sigma|$ obtained from 
\begin{equation}
 E_F = \tfrac{(\hbar k_F^\sigma)^2}{ 2m_{eff} } \pm \tfrac{1}{2}g^*_{xx}\mu_B B_x,
\end{equation}
where $\sigma=\uparrow,\downarrow$, the $\pm$ sign corresponds to spin down and up, respectively, and $D_c^{(n)}$ are extracted from Fig.~\ref{fig:onlyBz}, using Eq.~(\ref{eq:Dc}). Although the analytical lines are obtained neglecting the SOI and the Zeeman energy contribution from the transverse magnetic field, there is a good agreement between the obtained transport results and this simplified model. 

\begin{figure}[tb!]
 \includegraphics[width=0.99\columnwidth]{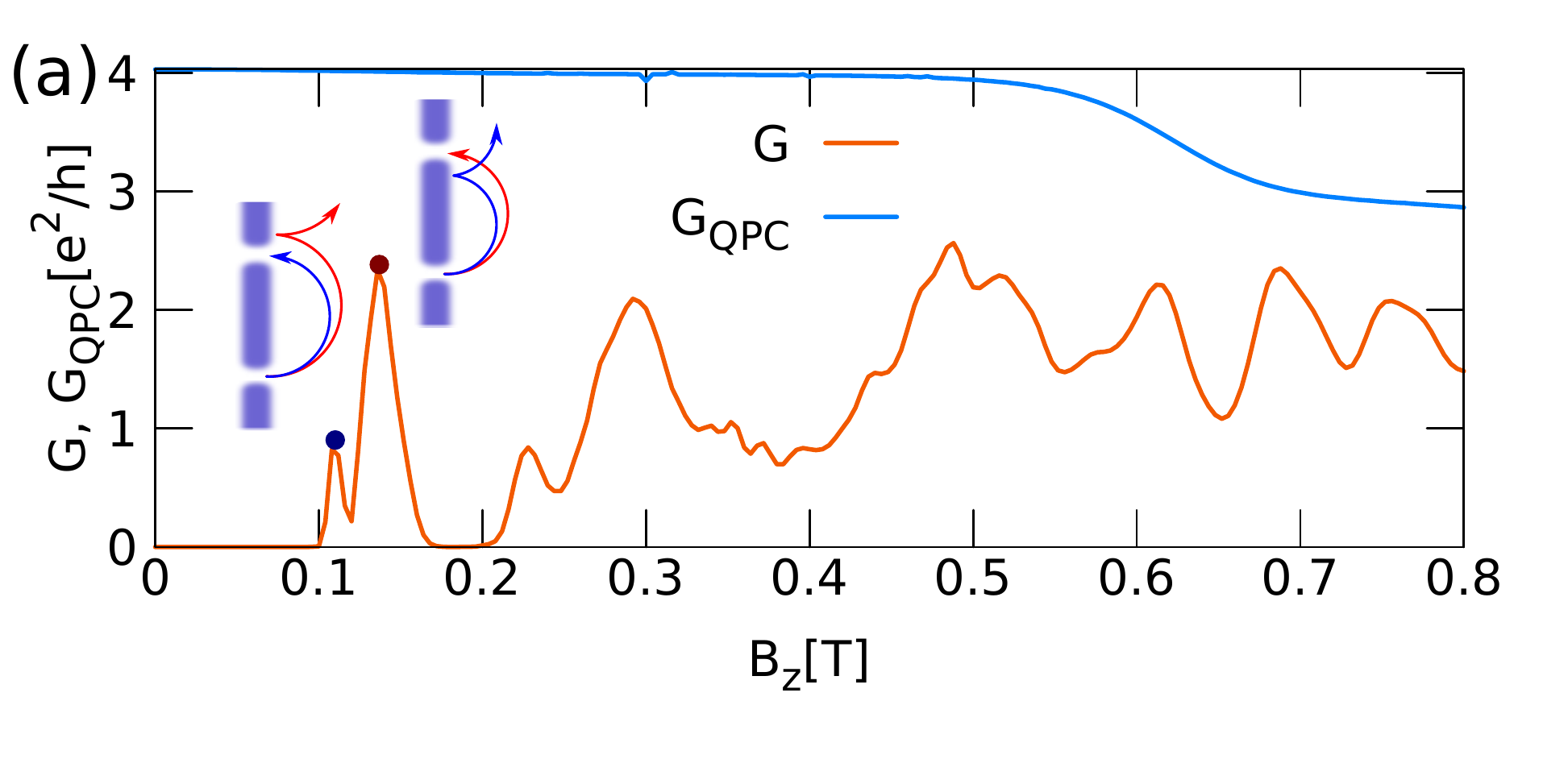}
 \includegraphics[width=0.99\columnwidth]{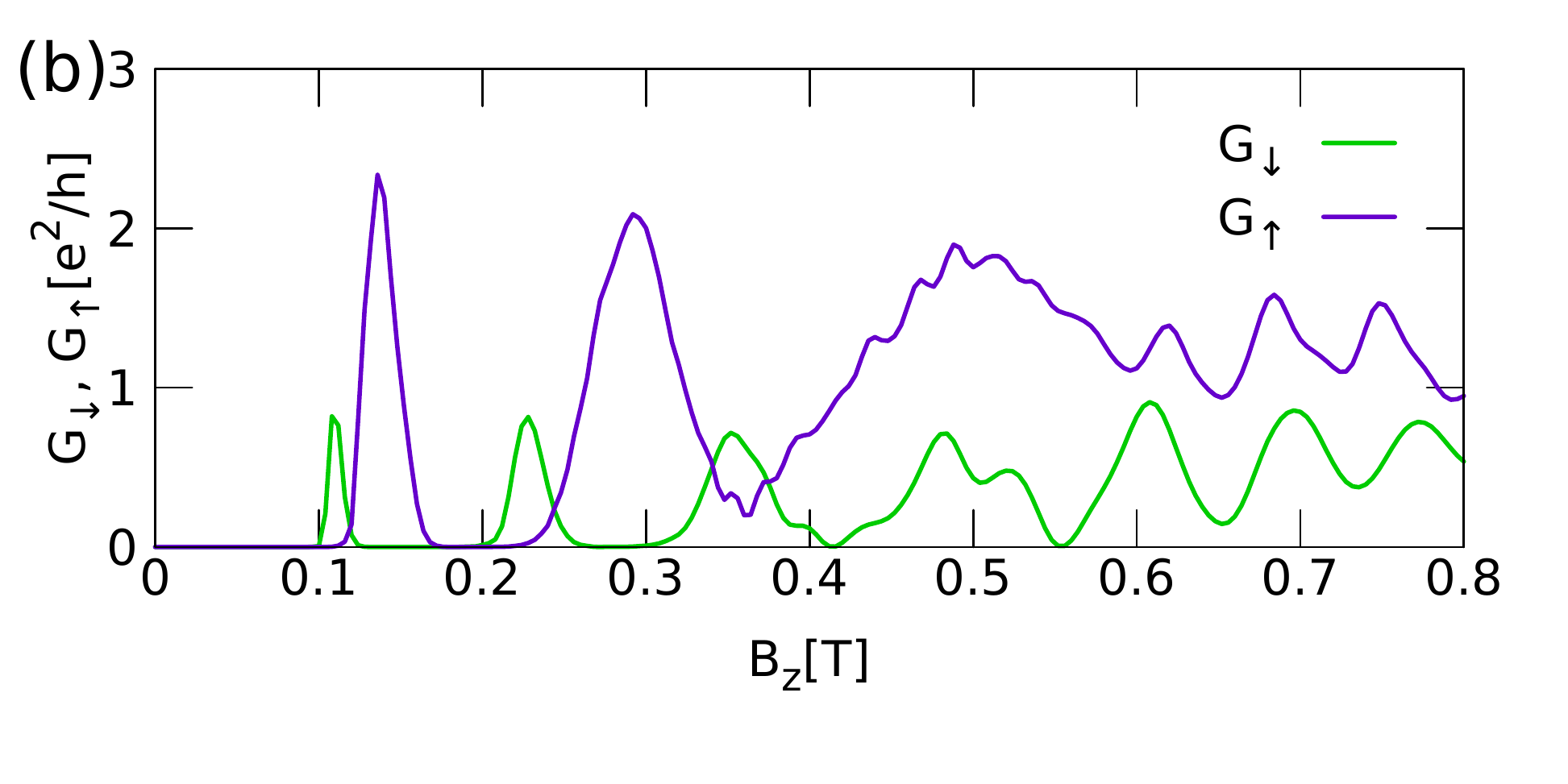}
  \caption{(a) The cross section of the conductance and the lower QPC conductance for $B_x=8$ T. (b) The spin resolved conductance.
   The first peak is split into two smaller peaks with $B_{z,\downarrow}^{(1)}=0.11$ T for spin down electrons and $B_{z,\uparrow}^{(1)}=0.137$ T for spin up electrons. 
   The inset in (a) shows semi-classical trajectories of the spin-up (red semi-circles) and spin-down (blue semi-circles) electron at the first focusing peak.
  } \label{fig:crossBx4}
\end{figure}

\begin{figure}[tb!]
 \includegraphics[width=0.334\columnwidth]{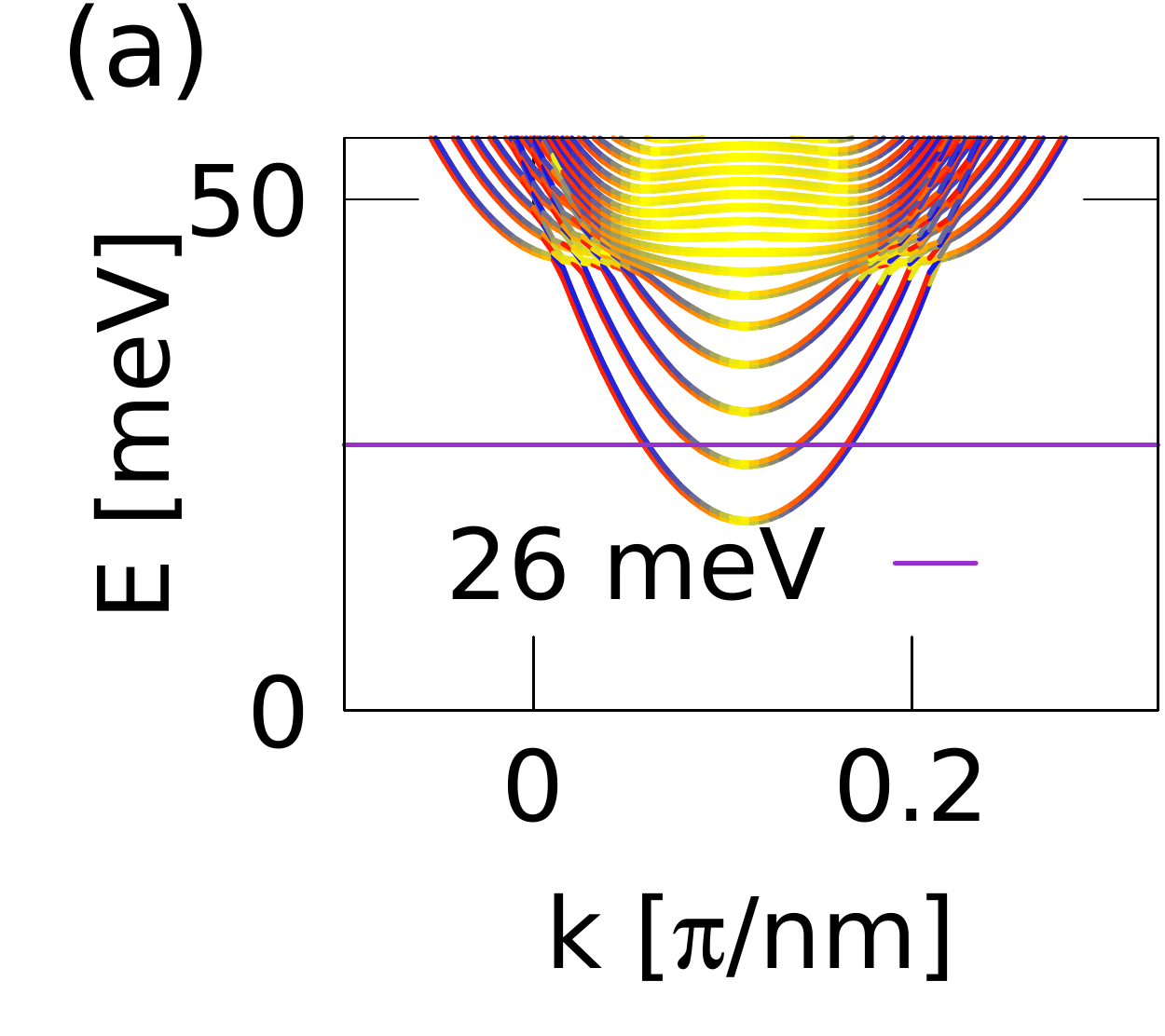}
 \includegraphics[width=0.283\columnwidth]{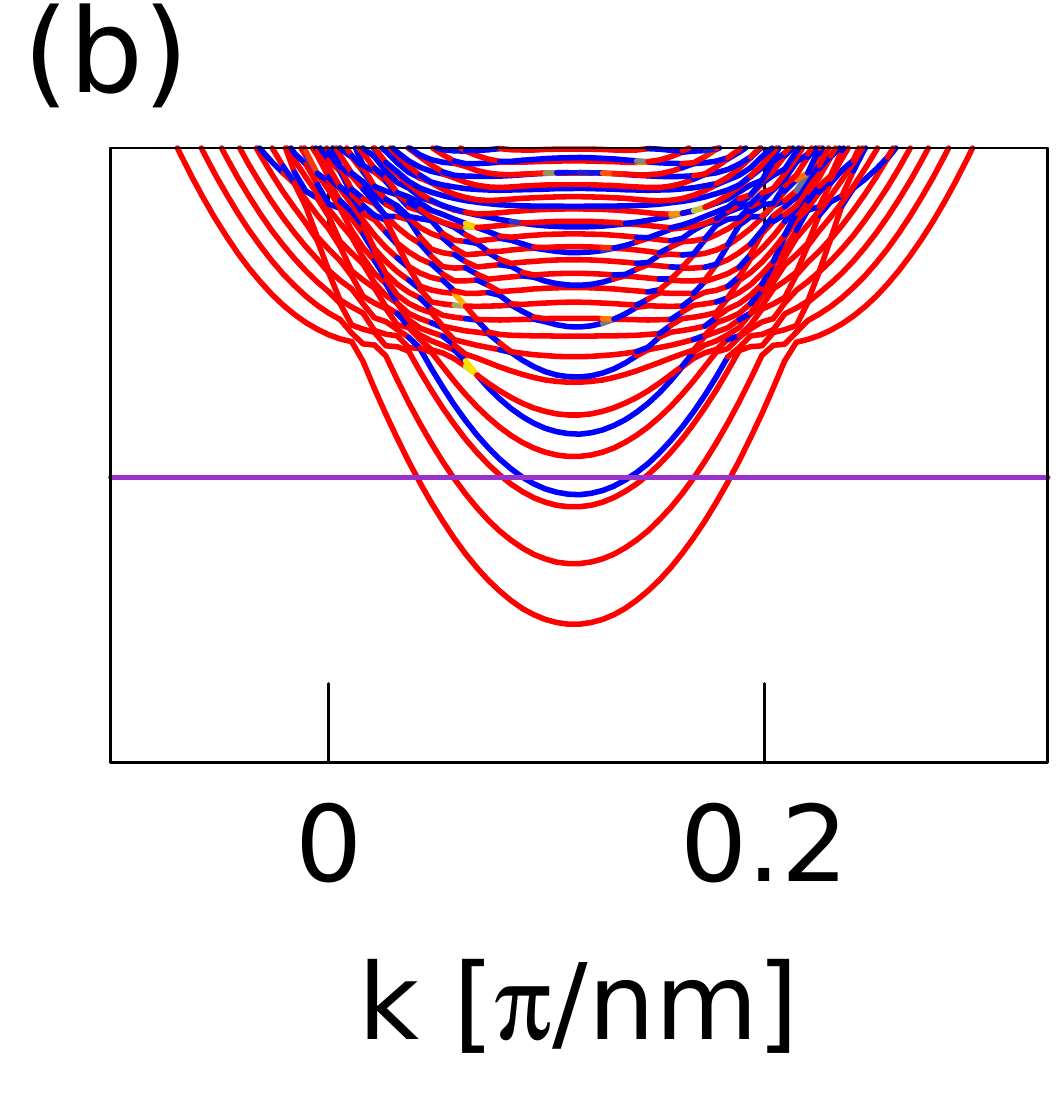}
 \includegraphics[width=0.358\columnwidth]{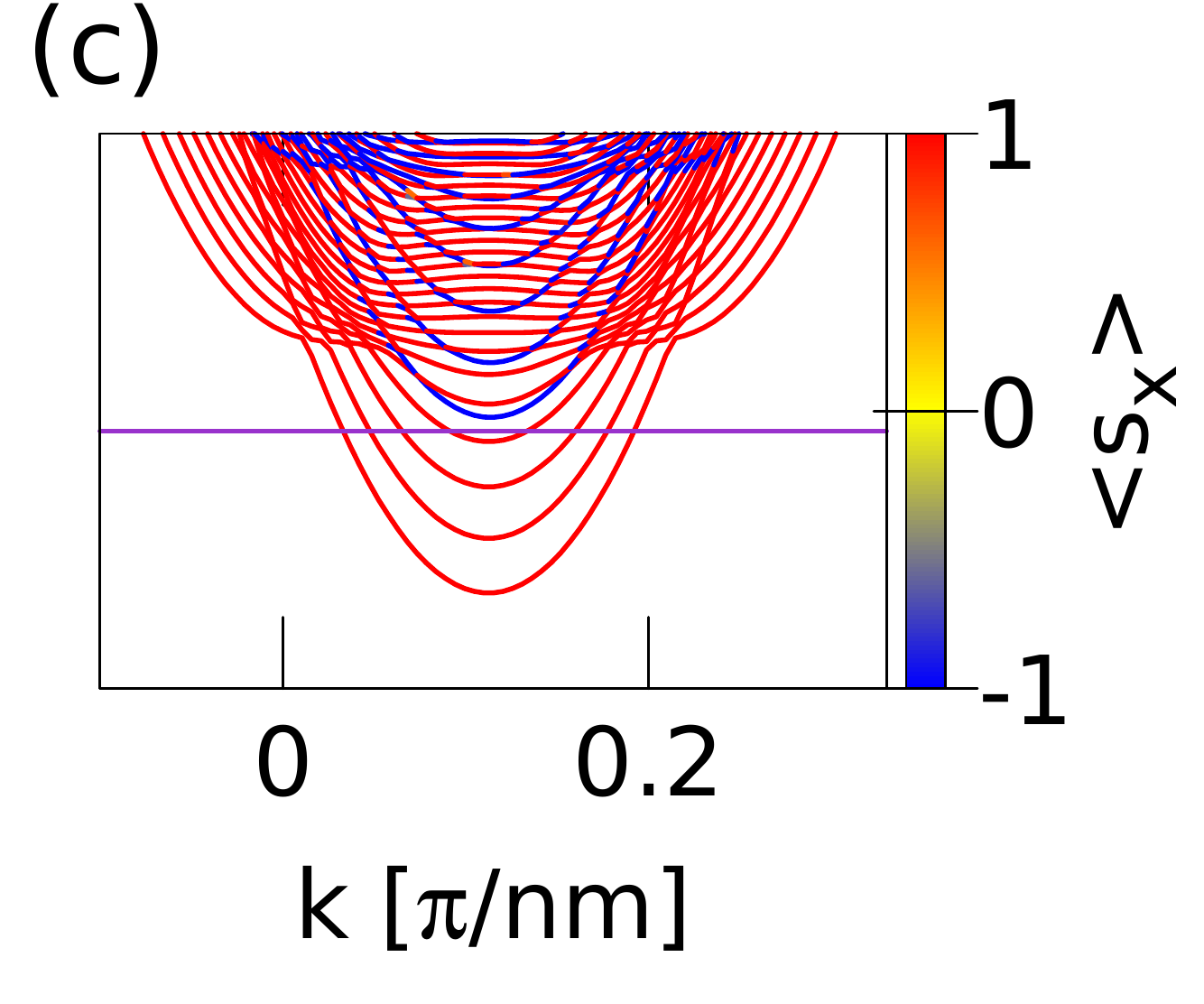}
  \caption{Dispersion relation of an infinite channel with the lateral potential taken
at the QPC constriction with $V_r=62$ meV, and (a) $B_x=0$ T (b) $B_x=8$ T and (c) 12 T. The color map shows the mean $x$ spin component of the subband. The spin-down subband shifts up in energy upon increasing $B_x$ and finally is raised above the Fermi energy. The opposite occurs for the spin-up electrons -- for increasing $B_x$ more and more subbands are available at the Fermi level.
  } \label{fig:dispQpcBx}
\end{figure}

 The cross section of the summed conductance and the spin-resolved conductance for $B_x=8$ T is shown in Fig.~\ref{fig:crossBx4}. 
In the pairs of focusing peaks, the spin-down (spin-up) conductance dominates
for the peak at lower (higher)  magnetic field [see Fig.~\ref{fig:crossBx4}(b)]. 
Interestingly, in each pair of the peaks in Fig.~\ref{fig:transpBx}, the lower one has smaller transmission than the upper one, and at $B_x\approx 10$ T vanishes, while the transmission of the upper one slowly increases. The reason for this behavior is the strong Zeeman splitting due to the in-plane magnetic field and the spin-dependent conductance of the QPCs \cite{Potok2002,Hanson2003}. 
Fig.~\ref{fig:dispQpcBx} shows the dispersion relation of an infinite channel with the lateral potential taken
at the QPC constriction with applied $B_x=0$, 8 T and 12 T. For $B_x=8$ T at the Fermi level for spin up 3 transverse subbands are available, while for spin down only one. For higher $B_x=12$ T the spin-down subband is raised above the Fermi level, and only spin-up electrons can pass through the QPC. On the other hand, for growing $B_x$, the number of spin-up subbands increases. Thus in the focusing spectrum, the upper peak -- the spin-up peak -- becomes more pronounced, while the lower one -- the spin-down peak -- has lower value of transmission and finally disappears.

\begin{figure}[tb!]
 \includegraphics[width=0.71\columnwidth]{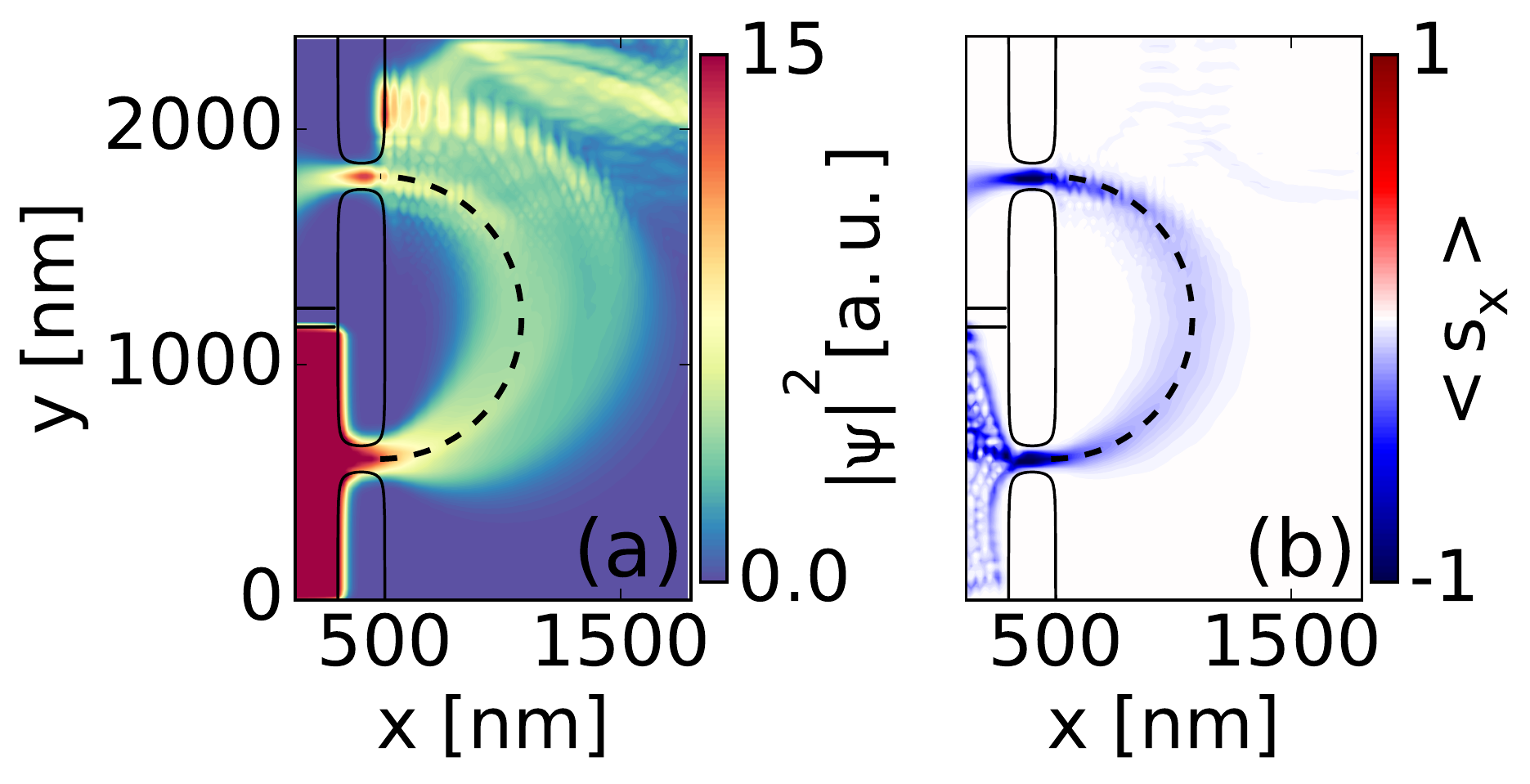}
 \includegraphics[width=0.273\columnwidth]{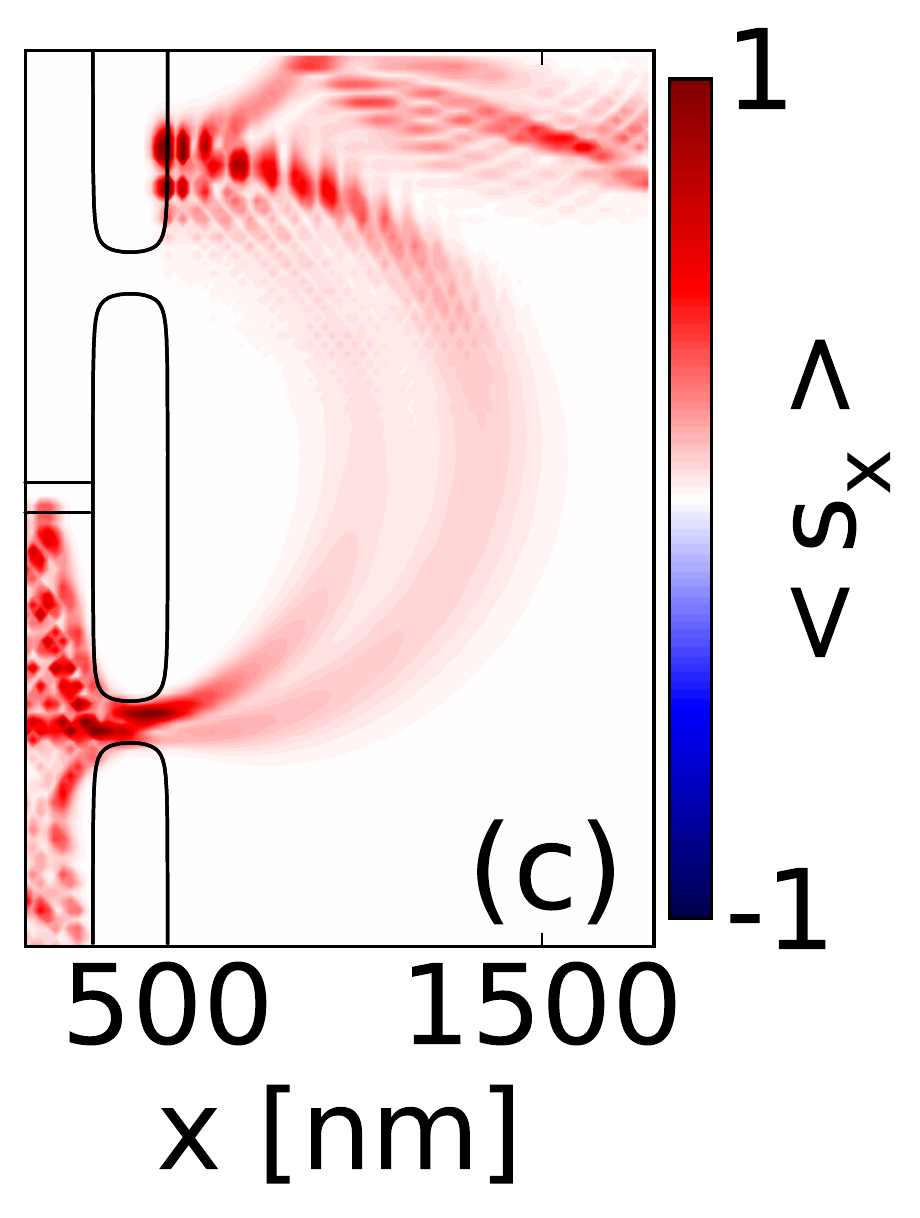}
  \caption{The density and average spins maps for the low-field peak in Fig.~\ref{fig:crossBx4}. In (b) the average spin $x$ projection for a spin-down mode is shown, and in (c) for a spin-up mode. The spin in the $y$ and $z$ directions is negligibly small (not shown), and the average spin in the $x$ direction is preserved (cf Fig. 13 for the spin precession effects in the case where SOI dominates over the Zeeman interaction).
  } \label{fig:densInSb}
\end{figure}

Concluding this section, we find that the in-plane magnetic field allows for a controllable separation of the electrons with opposite spins.
It is worth noting that in the systems that have strong SOI, without the in-plane magnetic field, only the odd focusing peaks get split \cite{Usaj2004, Reynoso2008, Lo2017}, and in case of the in-plane magnetic field all of the peaks are split. This is caused by the spin precession due to SOI in those systems. In our case the spin is determined by the effective magnetic field, which is almost parallel to $x$ direction. Thus the spin in $x$ direction dominates and the fluctuation due to SOI is negligible. It is shown in a representative case of the density and average spins for the low-field focusing peak at $B_{z,\downarrow}^{(1)}=0.11$ T in Fig.~\ref{fig:densInSb}. The electron spins are nearly unchanged along the entire path. The $\left\langle s_y \right\rangle$ and $\left\langle s_z \right\rangle$ are negligibly small compared to the $\left\langle s_x \right\rangle$.

\subsection{Scanning gate microscopy of the trajectories}

\begin{figure}[tb!]
\begin{center}
 \includegraphics[width=0.9\columnwidth]{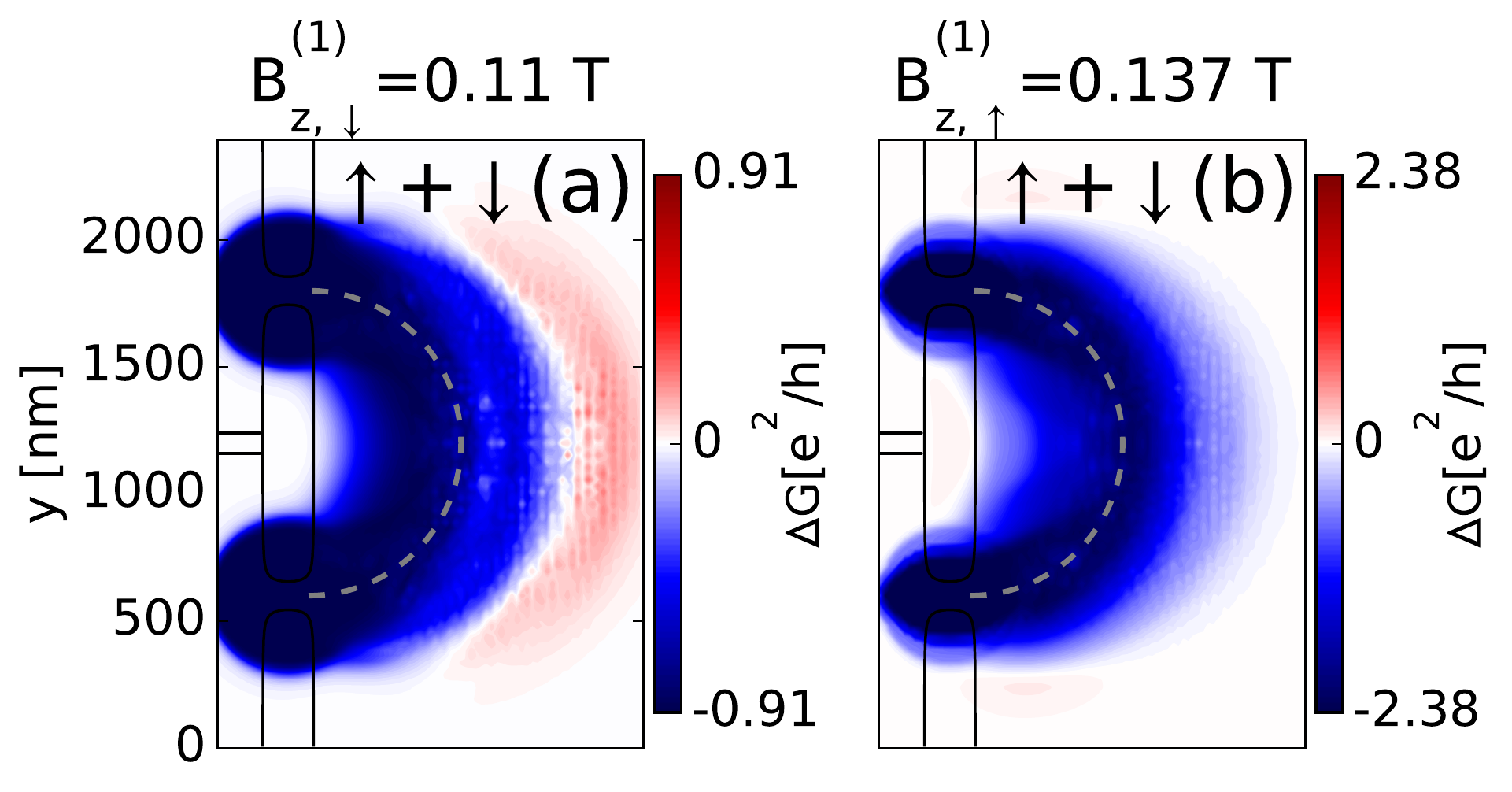}
 \includegraphics[width=0.9\columnwidth]{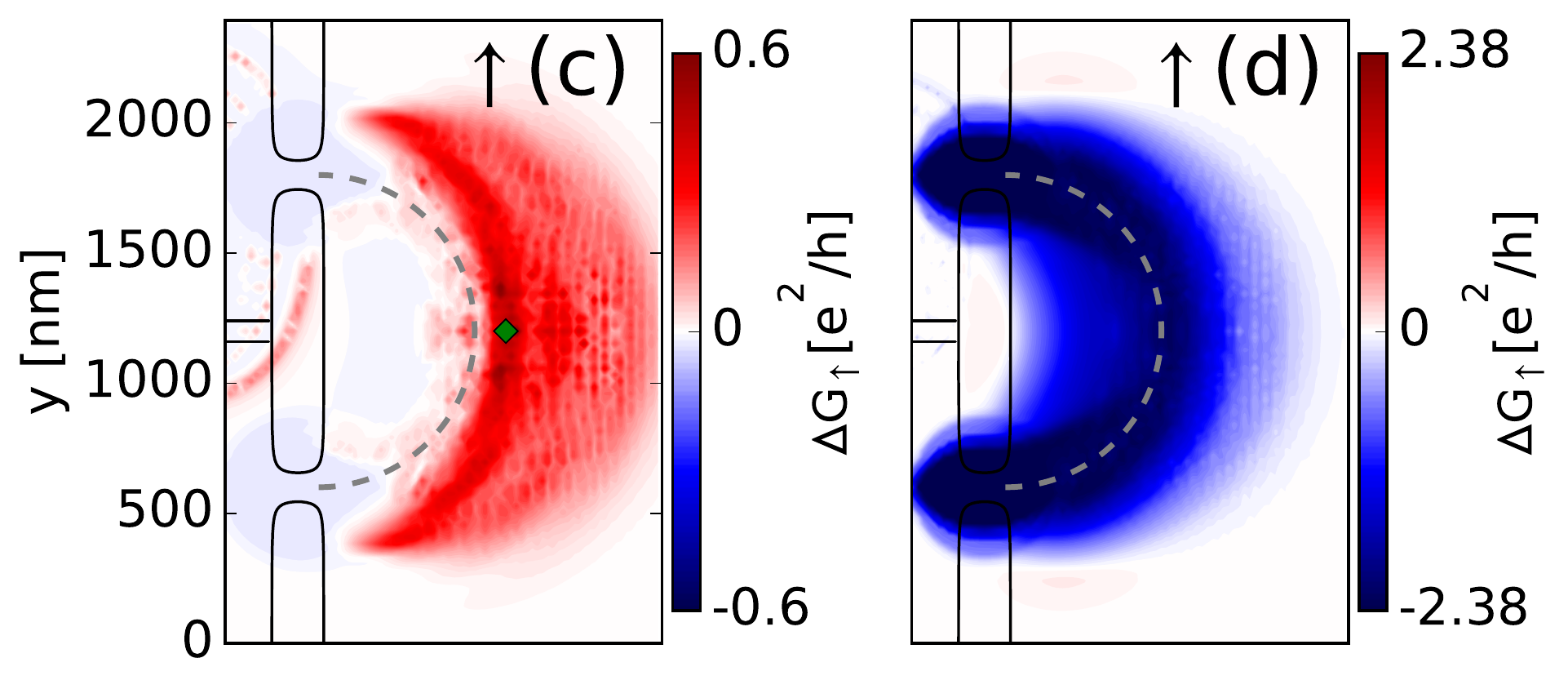}
 \includegraphics[width=0.9\columnwidth]{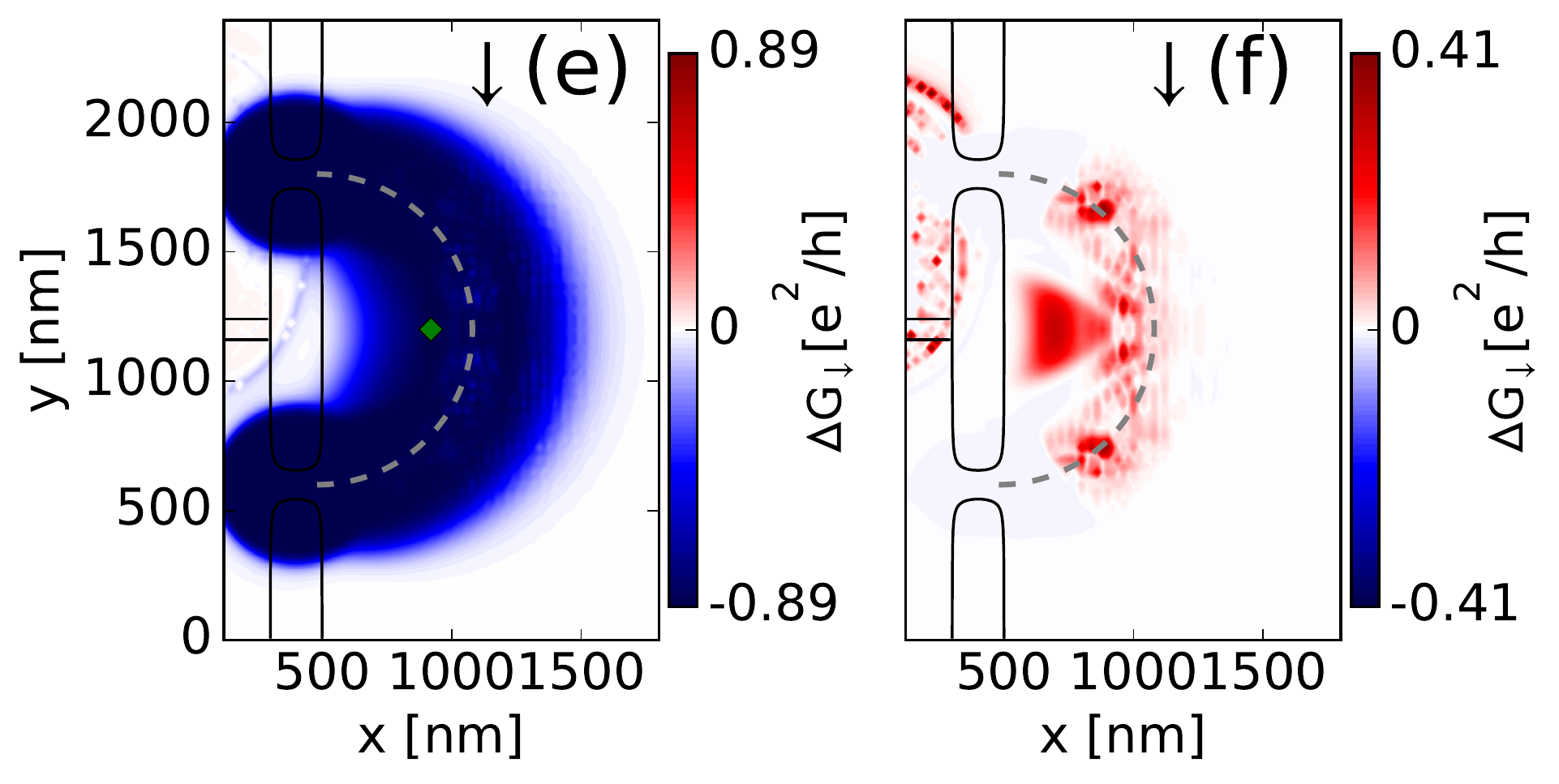}
\end{center}
  \caption{The conductance maps for the spin-down (left column) and spin-up (right column) focusing peak in Fig.~\ref{fig:crossBx4} at $B_{z,\downarrow}^{(1)}=0.11$ T and $B_{z,\uparrow}^{(1)}=0.137$ T, respectively. (a,b) the conductance summed over spins, (c,d) the spin-up conductance, (e,f) the spin-down conductance. The dashed semicircles show the semi-classical trajectory of an electron incident from the QPC with $k_x\ne 0$ only. The tiny arrows in the upper right corner show which contribution of $\Delta G$ is shown in the plot.
  } \label{fig:sgm1}
\end{figure}
\begin{figure}[tb!]
 \includegraphics[width=0.49\columnwidth]{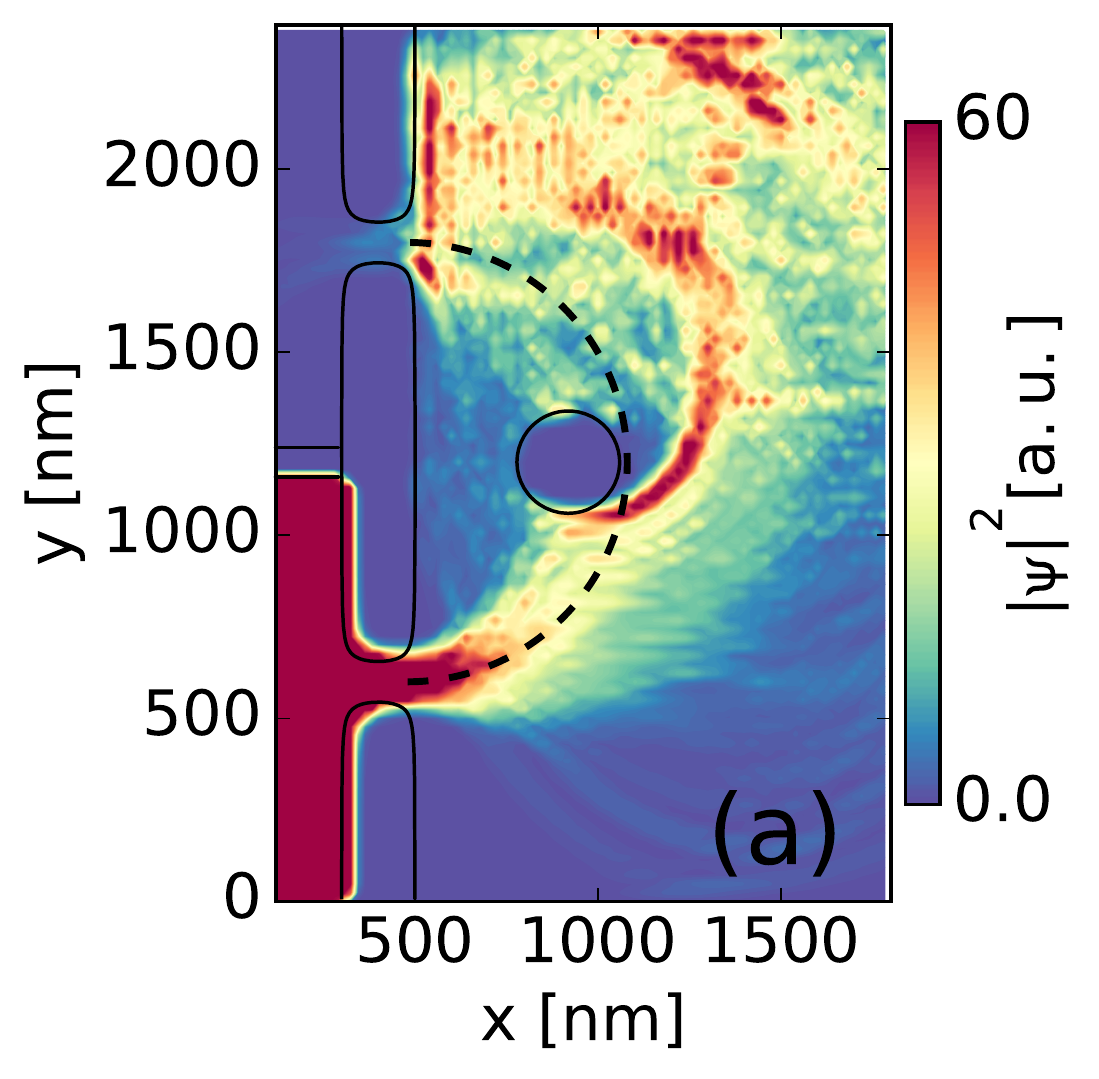}
 \includegraphics[width=0.49\columnwidth]{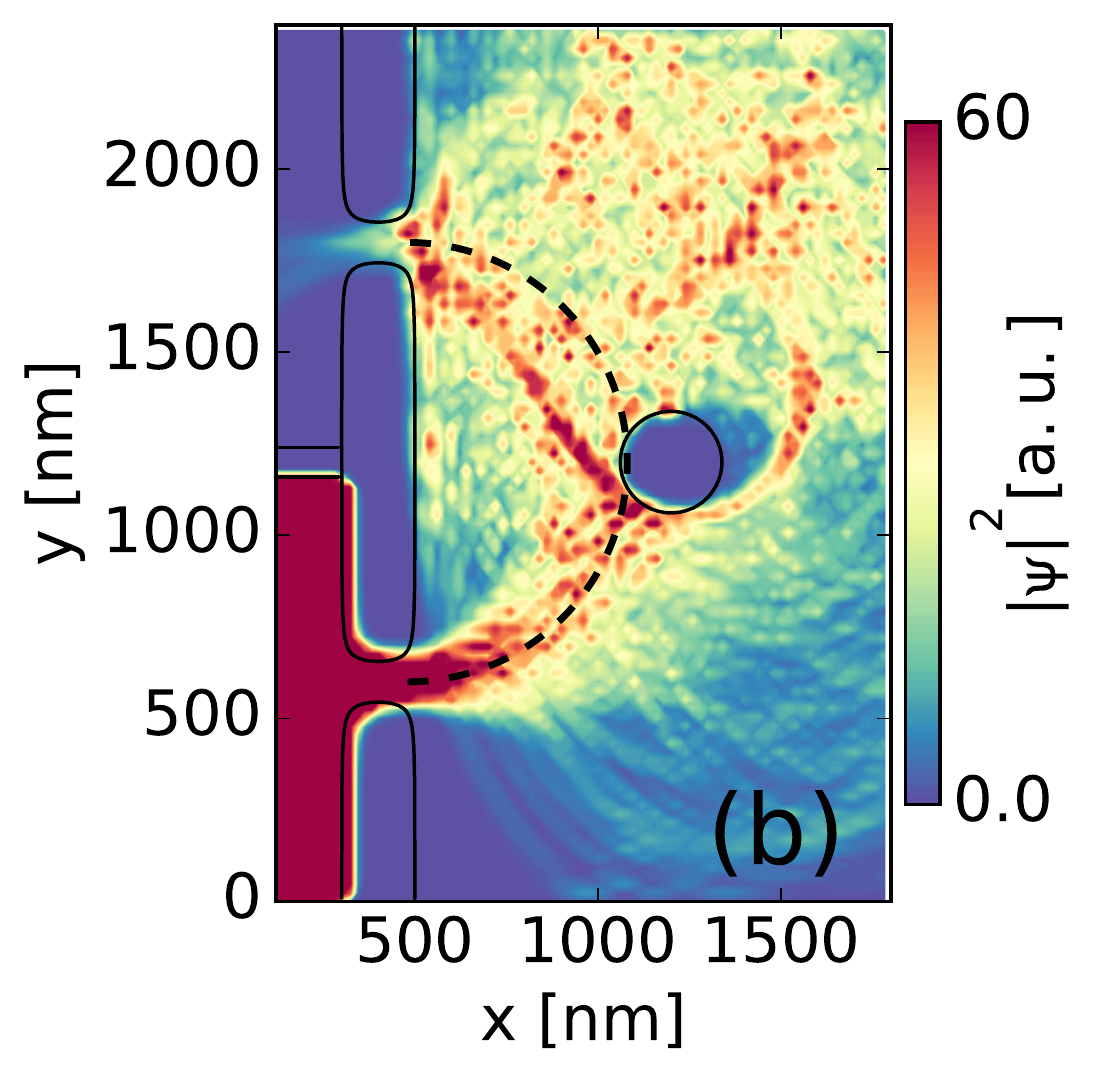}
  \caption{The density maps for the tip placed in the points marked with diamonds in Fig.~\ref{fig:sgm1}(c,e). (a) The tip blocking the beam with the tip at the point marked with green diamond in Fig.~\ref{fig:sgm1}(c). (b) The tip enabling the spin-up beam to enter the collector  with the tip at the point marked with green diamond in Fig.~\ref{fig:sgm1}(e).
  } \label{fig:densSGM}
\end{figure}

We simulated the SGM conductance maps for the magnetic fields that correspond to the peaks of magnetic focusing in the absence of the tip. We used $B_x=8$ T. In the cross section for $B_x=8$ T in Fig.~\ref{fig:crossBx4}(a) the dots show where the SGM scans were taken. Fig.~\ref{fig:sgm1} presents the maps of 
$\Delta G = G\left(\mathbf{r}_{tip}\right) - G\left(B_{z,\sigma}^{(1)}\right)$, and the spin-resolved conductances $\Delta G_{\sigma'} = G_{\sigma'}\left(\mathbf{r}_{tip}\right) - G_{\sigma'}\left(B_{z,\sigma}^{(1)}\right)$ with $\sigma,\sigma'=\uparrow,\downarrow$. The conductance maps exhibit semicircular pattern with a pronounced minimum along the semi-classical orbit of a carrier incident in the $x$ direction (indicated in Fig.~\ref{fig:sgm1} with dashed semi-circles). For the spin-up focusing peak at $B_{z,\downarrow}^{(1)}=0.11$ T, the scan [Fig.~\ref{fig:sgm1}(a)] is slightly different than for the spin-down peak at $B_{z,\uparrow}^{(1)}=0.137$ T [Fig.~\ref{fig:sgm1}(b)]. In the first one there is a slight increase of conductance to the right of the dashed semi-circle [see the red blob in Fig.~\ref{fig:sgm1}(a)]. Fig.~\ref{fig:sgm1}(c,d) show the spin-up conductance, and Fig.~\ref{fig:sgm1}(e,f) the spin-down conductance as a function of the tip position. One can see that in the spin-down peak (for $B_{z,\downarrow}^{(1)}=0.11$ T) the $\Delta G_{\downarrow}$ is everywhere negative or zero [Fig.~\ref{fig:sgm1}(e)], and $\Delta G_{\uparrow}$ -- positive or zero almost everywhere (except within the QPC) [Fig.~\ref{fig:sgm1}(c)]. Examples of electron densities with the tip placed in two different points are shown in Fig.~\ref{fig:densSGM}. In Fig.~\ref{fig:densSGM}(a), the tip, when placed along the electron trajectory leads to the deflection of the beam and blocks the spin-down beam, preventing it from entering the collector. 
On the other hand, in Fig.~\ref{fig:densSGM}(b), the tip can deflect the beam of spin-up electrons into the collector.

The situation is inverted in the peak at $B_z=0.137$ T. In the $\Delta G_{\uparrow}$ map [Fig.~\ref{fig:sgm1}(d)], the values are smaller or equal to zero, and in the $\Delta G_{\downarrow}$ map [Fig.~\ref{fig:sgm1}(f)], bigger or equal zero. In this case, the spin-up beam is blocked by the tip, thus $\Delta G_{\uparrow}$ drops along the semi-circle marked in Fig.~\ref{fig:sgm1}(d). On the other hand, the spin-down electrons have a smaller cyclotron diameter [than the QPC spacing $L$], but they can be scattered by the tip to the collector, which leads to an increase of $\Delta G$ at some points to the left of (or along) the dashed semi-circle.

\subsection{Magnetic focusing for heavy holes in GaAs/AlGaAs heterostructure }

We consider  an experiment conducted for two-dimensional hole gas (2DHG) in GaAs/AlGaAs, in Ref.~\onlinecite{Rokhinson2004}, where the splitting of the first focusing peak was visible without an in-plane magnetic field, and was solely due to the spin-orbit interaction. For this problem we assume the distance between the two QPCs $L=800$ nm, the computational box of width $W=1608$ nm and length $3000$ nm, the QPC defined in the same manner as in Eq.~\ref{eq:qpc_gates} with the geometrical parameters:  $l=500$ nm, $r=1100$ nm, $b_1=-600$ nm, $t_1=336$ nm, $t_2=468$ nm, $b_2=1140$ nm, $b_3=1272$ nm, $t_3=2208$ nm, and $d$=20 nm. We employ the effective mass of heavy holes $m_{eff}=0.17 m_e$ \cite{Plaut1988}, Land\'e factor $g_{zz}^*=-0.6$ \cite{Arora2013},
the Dresselhaus SO parameter $\beta=0.0477 $ eV{\AA } \cite{Rokhinson2004}, and zero Rashba SO. 

\begin{figure}[tb!]
 \includegraphics[width=0.99\columnwidth]{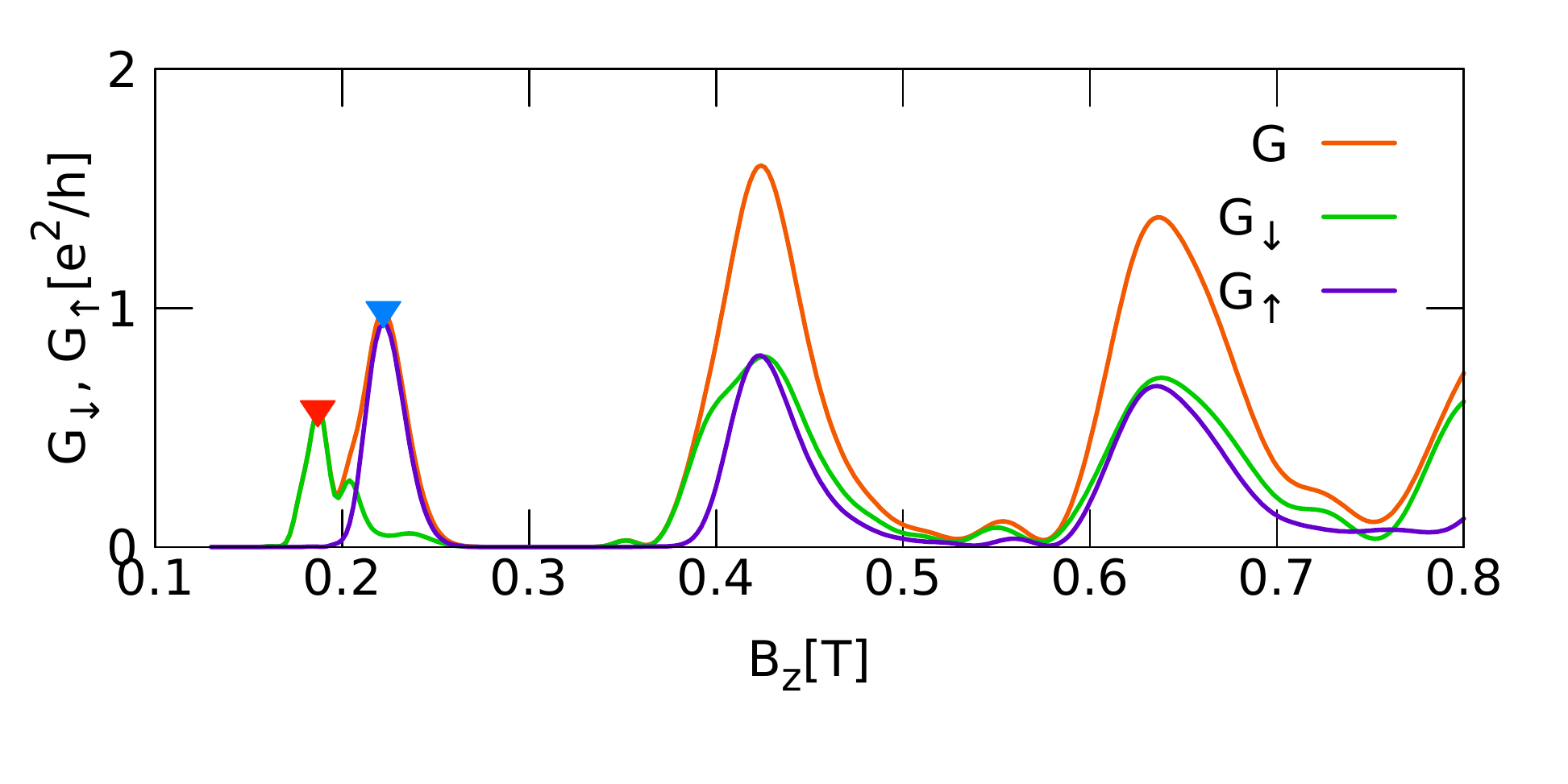}
  \caption{(a) The summed and spin resolved conductance for a hole system in GaAs/AlGaAs system. 
   The first peak is split into two smaller peaks with $B_{z,\downarrow}^{(1)}=0.187$ T for spin down holes and 
   $B_{z,\uparrow}^{(1)}=0.222$ T for spin up holes. 
    The peak splitting is 35 mT. 
  } \label{fig:crossHole}
\end{figure}

\begin{figure}[tb!]
 \includegraphics[width=0.5\columnwidth]{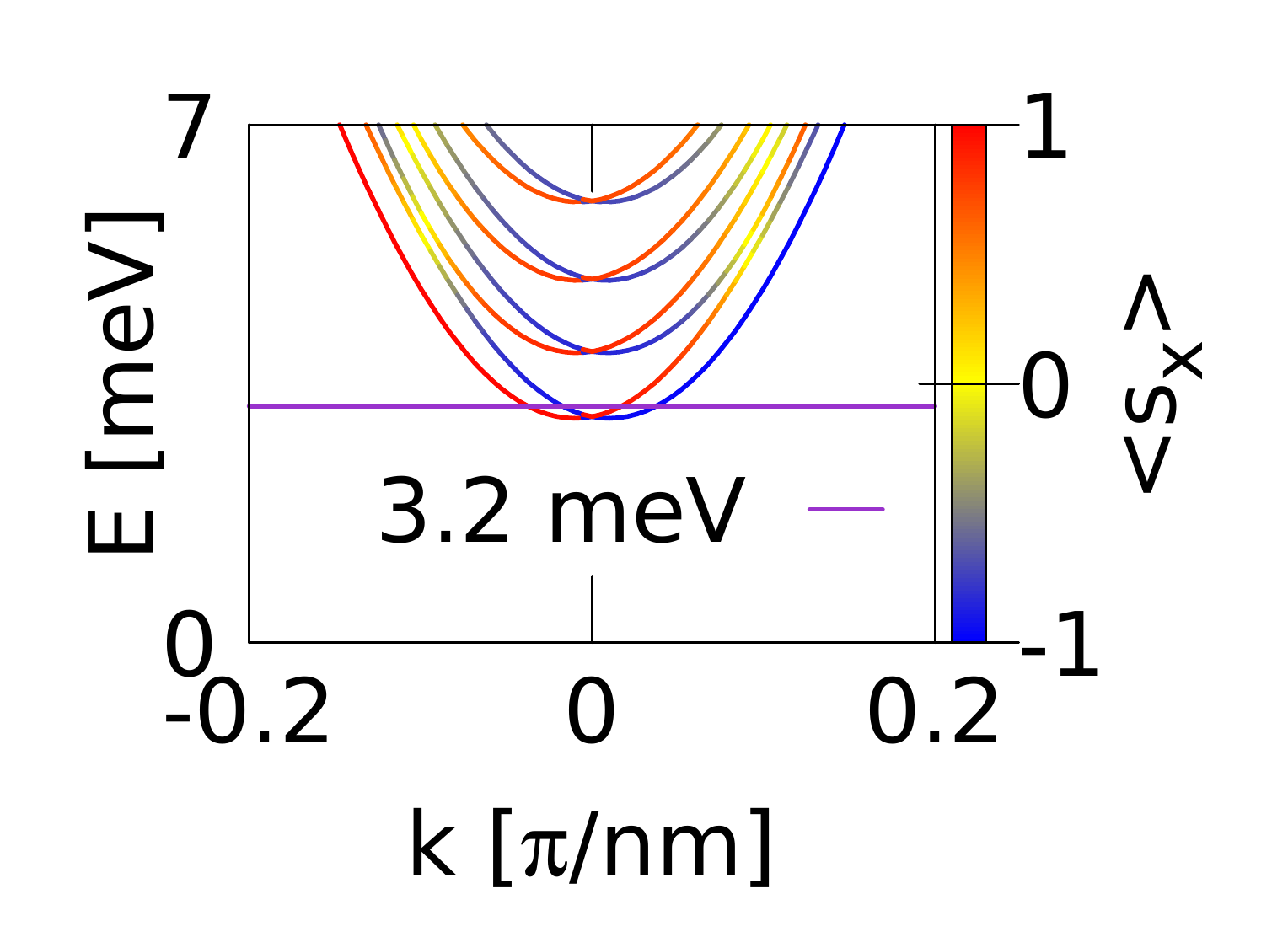}
  \caption{Dispersion relation of an infinite channel with the lateral potential taken
at the QPC constriction with $V_r=18$ meV, and $B_x=0$ T. The color map shows the mean $x$ spin component of the subbands. 
  } \label{fig:dispQpc_hole}
\end{figure}

We tune the lower QPC to $G_{QPC}=2e^2/h$, with $V_g=18$ meV, and $E_F=3.2$ meV. Figure \ref{fig:crossHole} shows the focusing conductance of the system. The focusing peaks are resolved, with the first peak split by 35 mT, remarkably close to the result in Ref.~\onlinecite{Rokhinson2004}, with the measured splitting of 36 mT. The splitting is due to the Dresselhaus SOI, which leads to the spin-polarization in the direction dependent on the hole momentum, and the difference in the Fermi wavenumbers $k_F$ of the holes with opposite spins. The band structure in the injector QPC is shown in Fig. ~\ref{fig:dispQpc_hole}. The hole spin in the injector QPC is in the $x$ direction. 

The difference in focusing magnetic field due to SOI can be evaluated by: 
\begin{equation}
 B_{z,\sigma}^{(1)} = \frac{2\hbar k_F^{\sigma}  }{ e D_c^{(1)} } = \frac{ \sqrt{2m_{eff} E_F} \mp m_{eff}\beta\hbar }{e D_c^{(1)} }.
\label{eq:dressBz}
\end{equation}

\begin{figure}[tb!]
 \includegraphics[width=0.71\columnwidth]{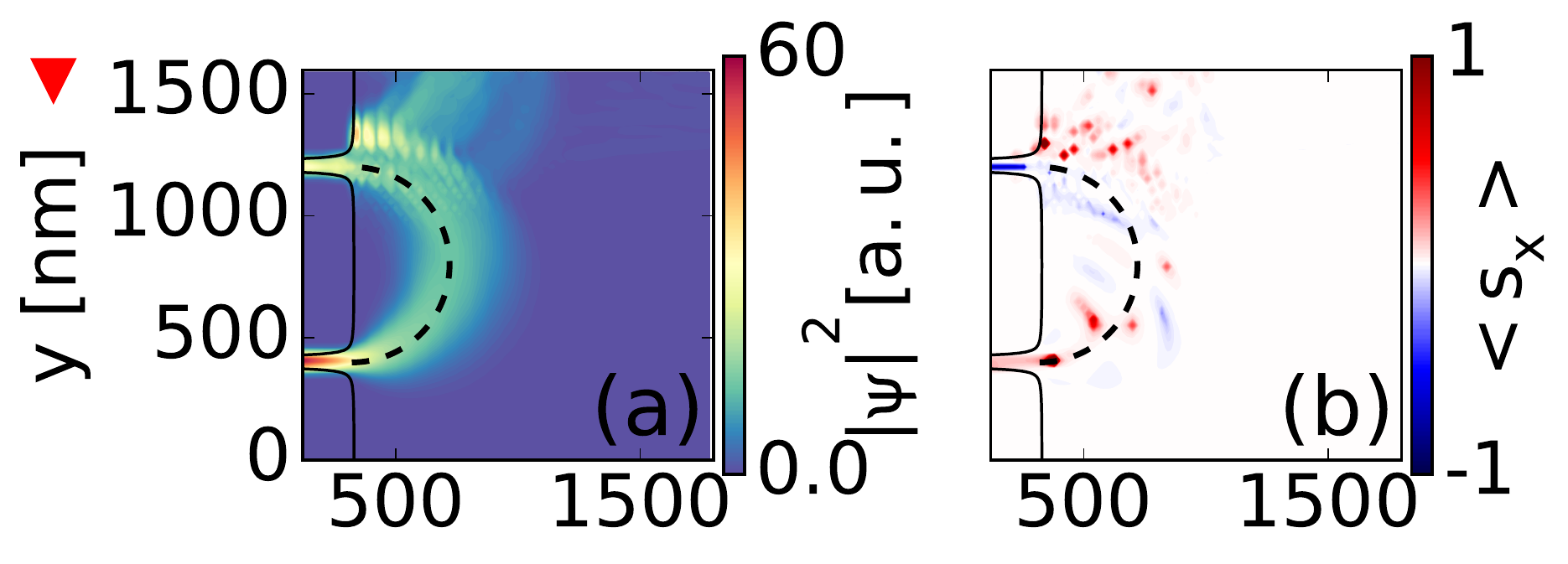}
 \includegraphics[width=0.275\columnwidth]{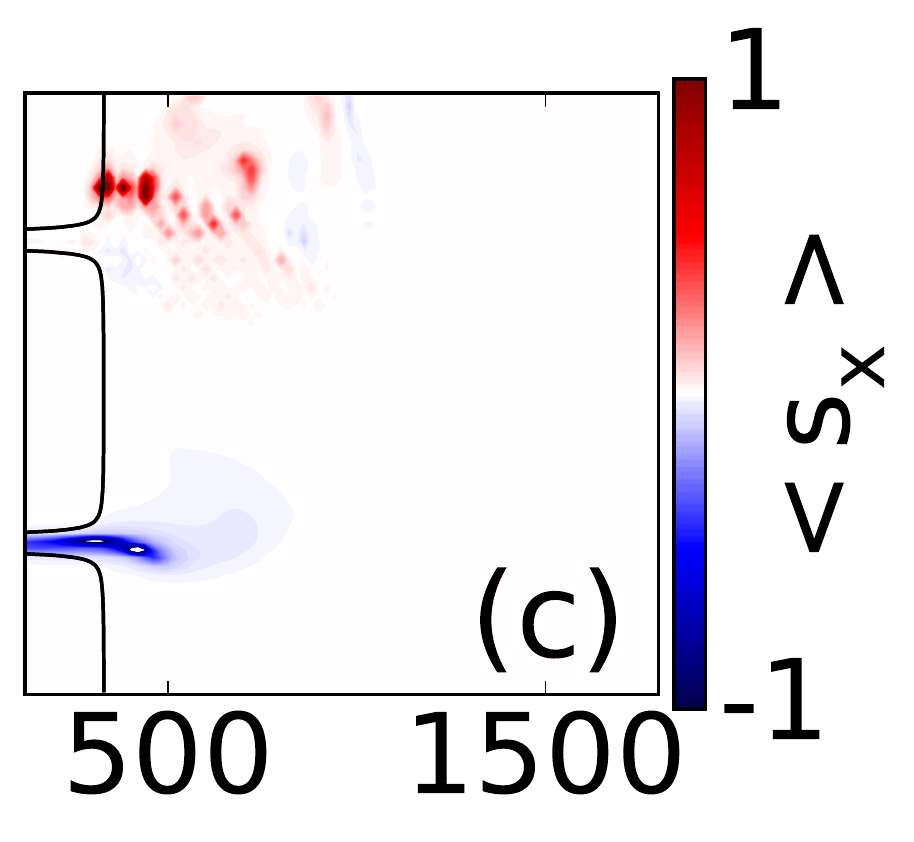}
 \includegraphics[width=0.71\columnwidth]{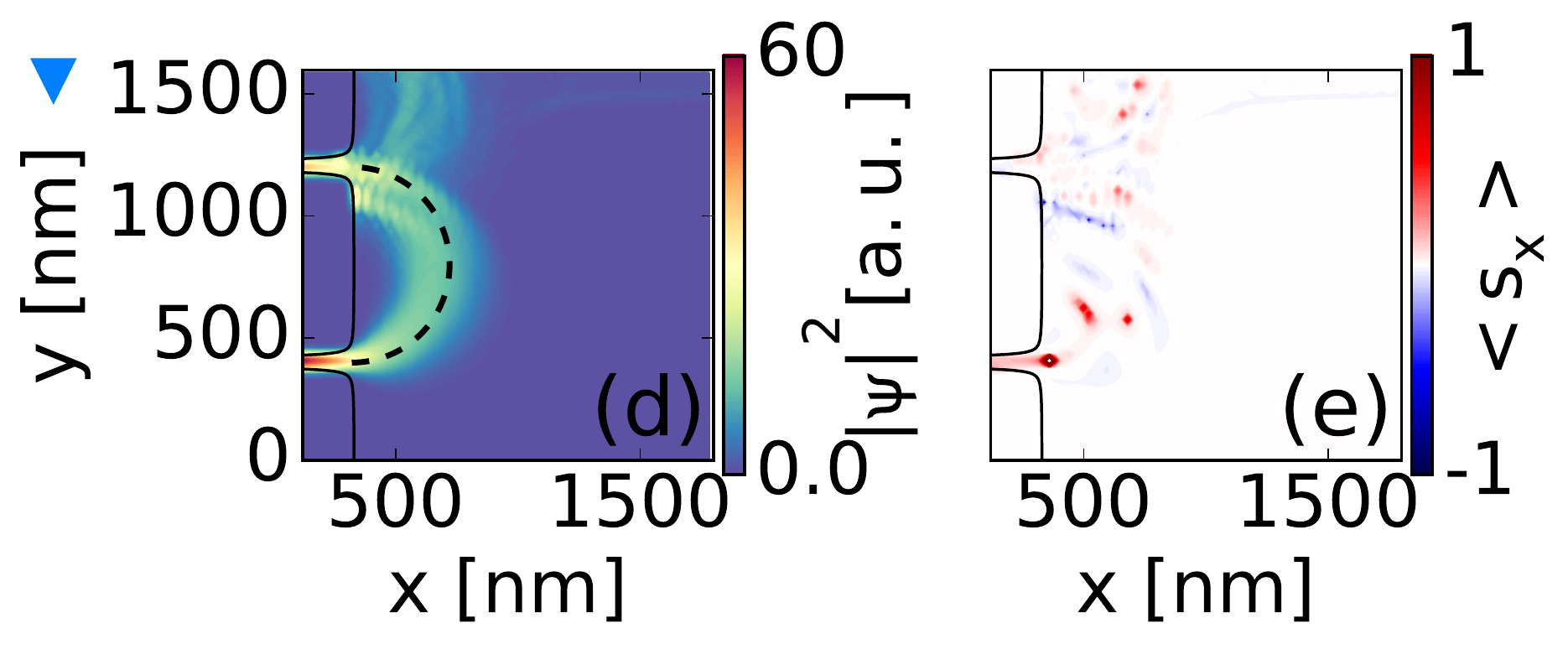}
 \includegraphics[width=0.275\columnwidth]{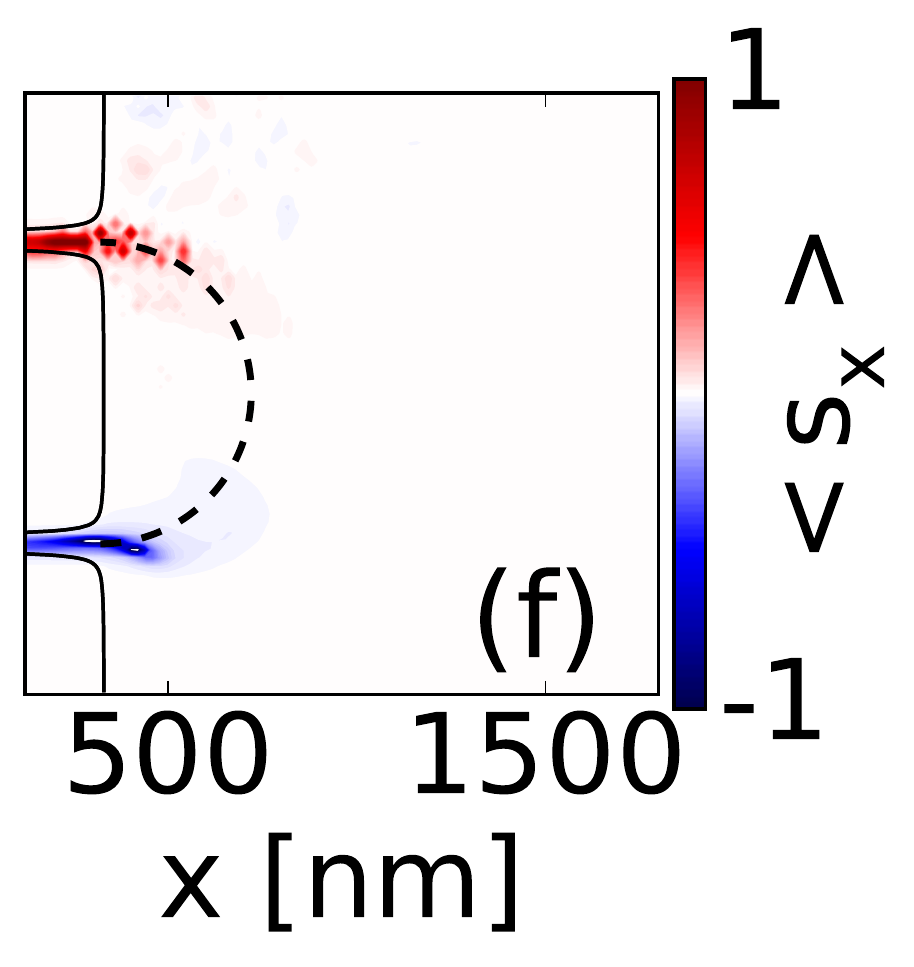}
  \caption{The density and average spin $x$ component maps for the focusing marked peaks in Fig.~\ref{fig:crossHole}, for the low-field peak marked with a red triangle (upper row) and high-field peak marked with a blue triangle (lower row). (a) and (b) The densities, (b) and (e) the average spin for the injected spin-up mode, and (c) and (f) for the injected spin-down mode. The flip of the spin direction in the detector QPC is visible. 
  } \label{fig:densHole}
\end{figure}

The density and the spin evolution in the peaks highlighted in Fig.~\ref{fig:crossHole} by tiny triangles is shown in Fig.~\ref{fig:densHole}. In the densities [Fig.~\ref{fig:densHole}(a,d)] the contributions of both spins with slightly different cyclotron radii are visible. In the averaged spin $x$ component maps for the mode injected with spin up [Fig.~\ref{fig:densHole}(b,e)] the precession is visible, but a little blurred due to the scattering from the gates' potential. For the mode injected with spin down [Fig.~\ref{fig:densHole}(c,f)] the flip of the spin direction in the detector is clearly visible.

\section{Summary and Conclusions}
We have studied the spatial spin-splitting of the electron trajectories
 in the transverse focusing system.
We demonstrated that the in-plane magnetic field of a few tesla in 
InSb induces the Zeeman splitting which is large enough to 
separate the conductance focusing peaks for the spin-down and spin-up
Fermi levels. The orientation of the spin is translated to the 
position of the conductance peak on the magnetic field scale. 
The focused trajectories for both the spin orientations can be resolved by
the scanning gate microscopy conductance maps. Moreover,
the SGM maps for opposite spin peaks contain qualitative differences 
due to the spin dependence of the cyclotron radii. 
The present finding 
paves the way for studies of the spin-dependent trajectories in the 
systems with the two-dimensional electron gas with high Land\'e factor materials. 

\section*{Acknowledgments}
This work was supported by the National Science Centre (NCN) Grant No. DEC-2015/17/N/ST3/02266,
and by AGH UST budget with the subsidy of the Ministry of
Science and Higher Education, Poland with Grant No. 15.11.220.718/6 for young researchers 
and Statutory Task No. 11.11.220.01/2.
The calculations were performed on PL-Grid and ACK CYFRONET AGH Infrastructure.

\bibliographystyle{apsrev4-1}
\bibliography{foc}

\begin{thebibliography}{48}%
\makeatletter
\providecommand \@ifxundefined [1]{%
 \@ifx{#1\undefined}
}%
\providecommand \@ifnum [1]{%
 \ifnum #1\expandafter \@firstoftwo
 \else \expandafter \@secondoftwo
 \fi
}%
\providecommand \@ifx [1]{%
 \ifx #1\expandafter \@firstoftwo
 \else \expandafter \@secondoftwo
 \fi
}%
\providecommand \natexlab [1]{#1}%
\providecommand \enquote  [1]{``#1''}%
\providecommand \bibnamefont  [1]{#1}%
\providecommand \bibfnamefont [1]{#1}%
\providecommand \citenamefont [1]{#1}%
\providecommand \href@noop [0]{\@secondoftwo}%
\providecommand \href [0]{\begingroup \@sanitize@url \@href}%
\providecommand \@href[1]{\@@startlink{#1}\@@href}%
\providecommand \@@href[1]{\endgroup#1\@@endlink}%
\providecommand \@sanitize@url [0]{\catcode `\\12\catcode `\$12\catcode
  `\&12\catcode `\#12\catcode `\^12\catcode `\_12\catcode `\%12\relax}%
\providecommand \@@startlink[1]{}%
\providecommand \@@endlink[0]{}%
\providecommand \url  [0]{\begingroup\@sanitize@url \@url }%
\providecommand \@url [1]{\endgroup\@href {#1}{\urlprefix }}%
\providecommand \urlprefix  [0]{URL }%
\providecommand \Eprint [0]{\href }%
\providecommand \doibase [0]{http://dx.doi.org/}%
\providecommand \selectlanguage [0]{\@gobble}%
\providecommand \bibinfo  [0]{\@secondoftwo}%
\providecommand \bibfield  [0]{\@secondoftwo}%
\providecommand \translation [1]{[#1]}%
\providecommand \BibitemOpen [0]{}%
\providecommand \bibitemStop [0]{}%
\providecommand \bibitemNoStop [0]{.\EOS\space}%
\providecommand \EOS [0]{\spacefactor3000\relax}%
\providecommand \BibitemShut  [1]{\csname bibitem#1\endcsname}%
\let\auto@bib@innerbib\@empty
\bibitem [{\citenamefont {Wolf}\ \emph {et~al.}(2001)\citenamefont {Wolf},
  \citenamefont {Awschalom}, \citenamefont {Buhrman}, \citenamefont {Daughton},
  \citenamefont {von Moln{\'a}r}, \citenamefont {Roukes}, \citenamefont
  {Chtchelkanova},\ and\ \citenamefont {Treger}}]{Wolf2001}%
  \BibitemOpen
  \bibfield  {author} {\bibinfo {author} {\bibfnamefont {S.~A.}\ \bibnamefont
  {Wolf}}, \bibinfo {author} {\bibfnamefont {D.~D.}\ \bibnamefont {Awschalom}},
  \bibinfo {author} {\bibfnamefont {R.~A.}\ \bibnamefont {Buhrman}}, \bibinfo
  {author} {\bibfnamefont {J.~M.}\ \bibnamefont {Daughton}}, \bibinfo {author}
  {\bibfnamefont {S.}~\bibnamefont {von Moln{\'a}r}}, \bibinfo {author}
  {\bibfnamefont {M.~L.}\ \bibnamefont {Roukes}}, \bibinfo {author}
  {\bibfnamefont {A.~Y.}\ \bibnamefont {Chtchelkanova}}, \ and\ \bibinfo
  {author} {\bibfnamefont {D.~M.}\ \bibnamefont {Treger}},\ }\href {\doibase
  10.1126/science.1065389} {\bibfield  {journal} {\bibinfo  {journal}
  {Science}\ }\textbf {\bibinfo {volume} {294}},\ \bibinfo {pages} {1488}
  (\bibinfo {year} {2001})}\BibitemShut {NoStop}%
\bibitem [{\citenamefont {Sharvin}(1965)}]{Sharvin1964}%
  \BibitemOpen
  \bibfield  {author} {\bibinfo {author} {\bibfnamefont {Y.~V.}\ \bibnamefont
  {Sharvin}},\ }\href@noop {} {\bibfield  {journal} {\bibinfo  {journal} {Sov.
  Phys. JETP}\ }\textbf {\bibinfo {volume} {21}},\ \bibinfo {pages} {655}
  (\bibinfo {year} {1965})}\BibitemShut {NoStop}%
\bibitem [{\citenamefont {Tsoi}(1974)}]{Tsoi1974}%
  \BibitemOpen
  \bibfield  {author} {\bibinfo {author} {\bibfnamefont {V.}~\bibnamefont
  {Tsoi}},\ }\href@noop {} {\bibfield  {journal} {\bibinfo  {journal} {JETP
  Lett.}\ }\textbf {\bibinfo {volume} {19}},\ \bibinfo {pages} {70} (\bibinfo
  {year} {1974})}\BibitemShut {NoStop}%
\bibitem [{\citenamefont {van Houten}\ \emph {et~al.}(1989)\citenamefont {van
  Houten}, \citenamefont {Beenakker}, \citenamefont {Williamson}, \citenamefont
  {Broekaart}, \citenamefont {van Loosdrecht}, \citenamefont {van Wees},
  \citenamefont {Mooij}, \citenamefont {Foxon},\ and\ \citenamefont
  {Harris}}]{Houten1989}%
  \BibitemOpen
  \bibfield  {author} {\bibinfo {author} {\bibfnamefont {H.}~\bibnamefont {van
  Houten}}, \bibinfo {author} {\bibfnamefont {C.~W.~J.}\ \bibnamefont
  {Beenakker}}, \bibinfo {author} {\bibfnamefont {J.~G.}\ \bibnamefont
  {Williamson}}, \bibinfo {author} {\bibfnamefont {M.~E.~I.}\ \bibnamefont
  {Broekaart}}, \bibinfo {author} {\bibfnamefont {P.~H.~M.}\ \bibnamefont {van
  Loosdrecht}}, \bibinfo {author} {\bibfnamefont {B.~J.}\ \bibnamefont {van
  Wees}}, \bibinfo {author} {\bibfnamefont {J.~E.}\ \bibnamefont {Mooij}},
  \bibinfo {author} {\bibfnamefont {C.~T.}\ \bibnamefont {Foxon}}, \ and\
  \bibinfo {author} {\bibfnamefont {J.~J.}\ \bibnamefont {Harris}},\ }\href
  {\doibase 10.1103/PhysRevB.39.8556} {\bibfield  {journal} {\bibinfo
  {journal} {Phys. Rev. B}\ }\textbf {\bibinfo {volume} {39}},\ \bibinfo
  {pages} {8556} (\bibinfo {year} {1989})}\BibitemShut {NoStop}%
\bibitem [{\citenamefont {Hanson}\ \emph {et~al.}(2003)\citenamefont {Hanson},
  \citenamefont {Witkamp}, \citenamefont {Vandersypen}, \citenamefont {van
  Beveren}, \citenamefont {Elzerman},\ and\ \citenamefont
  {Kouwenhoven}}]{Hanson2003}%
  \BibitemOpen
  \bibfield  {author} {\bibinfo {author} {\bibfnamefont {R.}~\bibnamefont
  {Hanson}}, \bibinfo {author} {\bibfnamefont {B.}~\bibnamefont {Witkamp}},
  \bibinfo {author} {\bibfnamefont {L.~M.~K.}\ \bibnamefont {Vandersypen}},
  \bibinfo {author} {\bibfnamefont {L.~H.~W.}\ \bibnamefont {van Beveren}},
  \bibinfo {author} {\bibfnamefont {J.~M.}\ \bibnamefont {Elzerman}}, \ and\
  \bibinfo {author} {\bibfnamefont {L.~P.}\ \bibnamefont {Kouwenhoven}},\
  }\href {\doibase 10.1103/PhysRevLett.91.196802} {\bibfield  {journal}
  {\bibinfo  {journal} {Phys. Rev. Lett.}\ }\textbf {\bibinfo {volume} {91}},\
  \bibinfo {pages} {196802} (\bibinfo {year} {2003})}\BibitemShut {NoStop}%
\bibitem [{\citenamefont {Aidala}\ \emph {et~al.}(2007)\citenamefont {Aidala},
  \citenamefont {Parrott}, \citenamefont {Kramer}, \citenamefont {Heller},
  \citenamefont {Westervelt}, \citenamefont {Hanson},\ and\ \citenamefont
  {Gossard}}]{Aidala2007}%
  \BibitemOpen
  \bibfield  {author} {\bibinfo {author} {\bibfnamefont {K.~E.}\ \bibnamefont
  {Aidala}}, \bibinfo {author} {\bibfnamefont {R.~E.}\ \bibnamefont {Parrott}},
  \bibinfo {author} {\bibfnamefont {T.}~\bibnamefont {Kramer}}, \bibinfo
  {author} {\bibfnamefont {E.~J.}\ \bibnamefont {Heller}}, \bibinfo {author}
  {\bibfnamefont {R.~M.}\ \bibnamefont {Westervelt}}, \bibinfo {author}
  {\bibfnamefont {M.~P.}\ \bibnamefont {Hanson}}, \ and\ \bibinfo {author}
  {\bibfnamefont {A.~C.}\ \bibnamefont {Gossard}},\ }\href
  {http://dx.doi.org/10.1038/nphys628} {\bibfield  {journal} {\bibinfo
  {journal} {Nat. Phys.}\ }\textbf {\bibinfo {volume} {3}},\ \bibinfo {pages}
  {464} (\bibinfo {year} {2007})}\BibitemShut {NoStop}%
\bibitem [{\citenamefont {Dedigama}\ \emph {et~al.}(2006)\citenamefont
  {Dedigama}, \citenamefont {Deen}, \citenamefont {Murphy}, \citenamefont
  {Goel}, \citenamefont {Keay}, \citenamefont {Santos}, \citenamefont {Suzuki},
  \citenamefont {Miyashita},\ and\ \citenamefont {Hirayama}}]{Dedigama2006}%
  \BibitemOpen
  \bibfield  {author} {\bibinfo {author} {\bibfnamefont {A.}~\bibnamefont
  {Dedigama}}, \bibinfo {author} {\bibfnamefont {D.}~\bibnamefont {Deen}},
  \bibinfo {author} {\bibfnamefont {S.}~\bibnamefont {Murphy}}, \bibinfo
  {author} {\bibfnamefont {N.}~\bibnamefont {Goel}}, \bibinfo {author}
  {\bibfnamefont {J.}~\bibnamefont {Keay}}, \bibinfo {author} {\bibfnamefont
  {M.}~\bibnamefont {Santos}}, \bibinfo {author} {\bibfnamefont
  {K.}~\bibnamefont {Suzuki}}, \bibinfo {author} {\bibfnamefont
  {S.}~\bibnamefont {Miyashita}}, \ and\ \bibinfo {author} {\bibfnamefont
  {Y.}~\bibnamefont {Hirayama}},\ }\href {\doibase
  https://doi.org/10.1016/j.physe.2006.03.050} {\bibfield  {journal} {\bibinfo
  {journal} {Physica E Low Dimens. Syst. Nanostruct.}\ }\textbf {\bibinfo
  {volume} {34}},\ \bibinfo {pages} {647} (\bibinfo {year} {2006})}\BibitemShut
  {NoStop}%
\bibitem [{\citenamefont {Lo}\ \emph {et~al.}(2017)\citenamefont {Lo},
  \citenamefont {Chen}, \citenamefont {Fan}, \citenamefont {Smith},
  \citenamefont {Creeth}, \citenamefont {Chang}, \citenamefont {Pepper},
  \citenamefont {Griffiths}, \citenamefont {Farrer}, \citenamefont {Beere},
  \citenamefont {Jones}, \citenamefont {Ritchie},\ and\ \citenamefont
  {Chen}}]{Lo2017}%
  \BibitemOpen
  \bibfield  {author} {\bibinfo {author} {\bibfnamefont {S.-T.}\ \bibnamefont
  {Lo}}, \bibinfo {author} {\bibfnamefont {C.-H.}\ \bibnamefont {Chen}},
  \bibinfo {author} {\bibfnamefont {J.-C.}\ \bibnamefont {Fan}}, \bibinfo
  {author} {\bibfnamefont {L.~W.}\ \bibnamefont {Smith}}, \bibinfo {author}
  {\bibfnamefont {G.~L.}\ \bibnamefont {Creeth}}, \bibinfo {author}
  {\bibfnamefont {C.-W.}\ \bibnamefont {Chang}}, \bibinfo {author}
  {\bibfnamefont {M.}~\bibnamefont {Pepper}}, \bibinfo {author} {\bibfnamefont
  {J.~P.}\ \bibnamefont {Griffiths}}, \bibinfo {author} {\bibfnamefont
  {I.}~\bibnamefont {Farrer}}, \bibinfo {author} {\bibfnamefont {H.~E.}\
  \bibnamefont {Beere}}, \bibinfo {author} {\bibfnamefont {G.~A.~C.}\
  \bibnamefont {Jones}}, \bibinfo {author} {\bibfnamefont {D.~A.}\ \bibnamefont
  {Ritchie}}, \ and\ \bibinfo {author} {\bibfnamefont {T.-M.}\ \bibnamefont
  {Chen}},\ }\href {http://dx.doi.org/10.1038/ncomms15997} {\bibfield
  {journal} {\bibinfo  {journal} {Nat. Commun.}\ }\textbf {\bibinfo {volume}
  {8}},\ \bibinfo {pages} {15997} (\bibinfo {year} {2017})}\BibitemShut
  {NoStop}%
\bibitem [{\citenamefont {Yan}\ \emph {et~al.}(2017)\citenamefont {Yan},
  \citenamefont {Kumar}, \citenamefont {Pepper}, \citenamefont {See},
  \citenamefont {Farrer}, \citenamefont {Ritchie}, \citenamefont {Griffiths},\
  and\ \citenamefont {Jones}}]{Yan2017}%
  \BibitemOpen
  \bibfield  {author} {\bibinfo {author} {\bibfnamefont {C.}~\bibnamefont
  {Yan}}, \bibinfo {author} {\bibfnamefont {S.}~\bibnamefont {Kumar}}, \bibinfo
  {author} {\bibfnamefont {M.}~\bibnamefont {Pepper}}, \bibinfo {author}
  {\bibfnamefont {P.}~\bibnamefont {See}}, \bibinfo {author} {\bibfnamefont
  {I.}~\bibnamefont {Farrer}}, \bibinfo {author} {\bibfnamefont
  {D.}~\bibnamefont {Ritchie}}, \bibinfo {author} {\bibfnamefont
  {J.}~\bibnamefont {Griffiths}}, \ and\ \bibinfo {author} {\bibfnamefont
  {G.}~\bibnamefont {Jones}},\ }\href {\doibase 10.1186/s11671-017-2321-4}
  {\bibfield  {journal} {\bibinfo  {journal} {Nanoscale Res. Lett.}\ }\textbf
  {\bibinfo {volume} {12}},\ \bibinfo {pages} {553} (\bibinfo {year}
  {2017})}\BibitemShut {NoStop}%
\bibitem [{\citenamefont {Rokhinson}\ \emph {et~al.}(2004)\citenamefont
  {Rokhinson}, \citenamefont {Larkina}, \citenamefont {Lyanda-Geller},
  \citenamefont {Pfeiffer},\ and\ \citenamefont {West}}]{Rokhinson2004}%
  \BibitemOpen
  \bibfield  {author} {\bibinfo {author} {\bibfnamefont {L.~P.}\ \bibnamefont
  {Rokhinson}}, \bibinfo {author} {\bibfnamefont {V.}~\bibnamefont {Larkina}},
  \bibinfo {author} {\bibfnamefont {Y.~B.}\ \bibnamefont {Lyanda-Geller}},
  \bibinfo {author} {\bibfnamefont {L.~N.}\ \bibnamefont {Pfeiffer}}, \ and\
  \bibinfo {author} {\bibfnamefont {K.~W.}\ \bibnamefont {West}},\ }\href
  {\doibase 10.1103/PhysRevLett.93.146601} {\bibfield  {journal} {\bibinfo
  {journal} {Phys. Rev. Lett.}\ }\textbf {\bibinfo {volume} {93}},\ \bibinfo
  {pages} {146601} (\bibinfo {year} {2004})}\BibitemShut {NoStop}%
\bibitem [{\citenamefont {Chesi}\ \emph {et~al.}(2011)\citenamefont {Chesi},
  \citenamefont {Giuliani}, \citenamefont {Rokhinson}, \citenamefont
  {Pfeiffer},\ and\ \citenamefont {West}}]{Chesi2011}%
  \BibitemOpen
  \bibfield  {author} {\bibinfo {author} {\bibfnamefont {S.}~\bibnamefont
  {Chesi}}, \bibinfo {author} {\bibfnamefont {G.~F.}\ \bibnamefont {Giuliani}},
  \bibinfo {author} {\bibfnamefont {L.~P.}\ \bibnamefont {Rokhinson}}, \bibinfo
  {author} {\bibfnamefont {L.~N.}\ \bibnamefont {Pfeiffer}}, \ and\ \bibinfo
  {author} {\bibfnamefont {K.~W.}\ \bibnamefont {West}},\ }\href {\doibase
  10.1103/PhysRevLett.106.236601} {\bibfield  {journal} {\bibinfo  {journal}
  {Phys. Rev. Lett.}\ }\textbf {\bibinfo {volume} {106}},\ \bibinfo {pages}
  {236601} (\bibinfo {year} {2011})}\BibitemShut {NoStop}%
\bibitem [{\citenamefont {Rokhinson}\ \emph {et~al.}(2006)\citenamefont
  {Rokhinson}, \citenamefont {Pfeiffer},\ and\ \citenamefont
  {West}}]{Rokhinson2006}%
  \BibitemOpen
  \bibfield  {author} {\bibinfo {author} {\bibfnamefont {L.~P.}\ \bibnamefont
  {Rokhinson}}, \bibinfo {author} {\bibfnamefont {L.~N.}\ \bibnamefont
  {Pfeiffer}}, \ and\ \bibinfo {author} {\bibfnamefont {K.~W.}\ \bibnamefont
  {West}},\ }\href {\doibase 10.1103/PhysRevLett.96.156602} {\bibfield
  {journal} {\bibinfo  {journal} {Phys. Rev. Lett.}\ }\textbf {\bibinfo
  {volume} {96}},\ \bibinfo {pages} {156602} (\bibinfo {year}
  {2006})}\BibitemShut {NoStop}%
\bibitem [{\citenamefont {Sellier}\ \emph {et~al.}(2011)\citenamefont
  {Sellier}, \citenamefont {Hackens}, \citenamefont {Pala}, \citenamefont
  {Martins}, \citenamefont {Baltazar}, \citenamefont {Wallart}, \citenamefont
  {Desplanque}, \citenamefont {Bayot},\ and\ \citenamefont
  {Huant}}]{Sellier2011}%
  \BibitemOpen
  \bibfield  {author} {\bibinfo {author} {\bibfnamefont {H.}~\bibnamefont
  {Sellier}}, \bibinfo {author} {\bibfnamefont {B.}~\bibnamefont {Hackens}},
  \bibinfo {author} {\bibfnamefont {M.~G.}\ \bibnamefont {Pala}}, \bibinfo
  {author} {\bibfnamefont {F.}~\bibnamefont {Martins}}, \bibinfo {author}
  {\bibfnamefont {S.}~\bibnamefont {Baltazar}}, \bibinfo {author}
  {\bibfnamefont {X.}~\bibnamefont {Wallart}}, \bibinfo {author} {\bibfnamefont
  {L.}~\bibnamefont {Desplanque}}, \bibinfo {author} {\bibfnamefont
  {V.}~\bibnamefont {Bayot}}, \ and\ \bibinfo {author} {\bibfnamefont
  {S.}~\bibnamefont {Huant}},\ }\href
  {http://stacks.iop.org/0268-1242/26/i=6/a=064008} {\bibfield  {journal}
  {\bibinfo  {journal} {Semicond. Sci. Technol.}\ }\textbf {\bibinfo {volume}
  {26}},\ \bibinfo {pages} {064008} (\bibinfo {year} {2011})}\BibitemShut
  {NoStop}%
\bibitem [{\citenamefont {Usaj}\ and\ \citenamefont
  {Balseiro}(2004)}]{Usaj2004}%
  \BibitemOpen
  \bibfield  {author} {\bibinfo {author} {\bibfnamefont {G.}~\bibnamefont
  {Usaj}}\ and\ \bibinfo {author} {\bibfnamefont {C.~A.}\ \bibnamefont
  {Balseiro}},\ }\href {\doibase 10.1103/PhysRevB.70.041301} {\bibfield
  {journal} {\bibinfo  {journal} {Phys. Rev. B}\ }\textbf {\bibinfo {volume}
  {70}},\ \bibinfo {pages} {041301} (\bibinfo {year} {2004})}\BibitemShut
  {NoStop}%
\bibitem [{\citenamefont {Z\"ulicke}\ \emph {et~al.}(2007)\citenamefont
  {Z\"ulicke}, \citenamefont {Bolte},\ and\ \citenamefont
  {Winkler}}]{Zulicke2007}%
  \BibitemOpen
  \bibfield  {author} {\bibinfo {author} {\bibfnamefont {U.}~\bibnamefont
  {Z\"ulicke}}, \bibinfo {author} {\bibfnamefont {J.}~\bibnamefont {Bolte}}, \
  and\ \bibinfo {author} {\bibfnamefont {R.}~\bibnamefont {Winkler}},\ }\href
  {http://stacks.iop.org/1367-2630/9/i=9/a=355} {\bibfield  {journal} {\bibinfo
   {journal} {New Journal of Physics}\ }\textbf {\bibinfo {volume} {9}},\
  \bibinfo {pages} {355} (\bibinfo {year} {2007})}\BibitemShut {NoStop}%
\bibitem [{\citenamefont {Reynoso}\ \emph {et~al.}(2007)\citenamefont
  {Reynoso}, \citenamefont {Usaj},\ and\ \citenamefont
  {Balseiro}}]{Reynoso2007}%
  \BibitemOpen
  \bibfield  {author} {\bibinfo {author} {\bibfnamefont {A.}~\bibnamefont
  {Reynoso}}, \bibinfo {author} {\bibfnamefont {G.}~\bibnamefont {Usaj}}, \
  and\ \bibinfo {author} {\bibfnamefont {C.~A.}\ \bibnamefont {Balseiro}},\
  }\href {\doibase 10.1103/PhysRevB.75.085321} {\bibfield  {journal} {\bibinfo
  {journal} {Phys. Rev. B}\ }\textbf {\bibinfo {volume} {75}},\ \bibinfo
  {pages} {085321} (\bibinfo {year} {2007})}\BibitemShut {NoStop}%
\bibitem [{\citenamefont {Schliemann}(2008)}]{Schliemann2008}%
  \BibitemOpen
  \bibfield  {author} {\bibinfo {author} {\bibfnamefont {J.}~\bibnamefont
  {Schliemann}},\ }\href {\doibase 10.1103/PhysRevB.77.125303} {\bibfield
  {journal} {\bibinfo  {journal} {Phys. Rev. B}\ }\textbf {\bibinfo {volume}
  {77}},\ \bibinfo {pages} {125303} (\bibinfo {year} {2008})}\BibitemShut
  {NoStop}%
\bibitem [{\citenamefont {Reynoso}\ \emph
  {et~al.}(2008{\natexlab{a}})\citenamefont {Reynoso}, \citenamefont {Usaj},\
  and\ \citenamefont {Balseiro}}]{Reynoso2008b}%
  \BibitemOpen
  \bibfield  {author} {\bibinfo {author} {\bibfnamefont {A.~A.}\ \bibnamefont
  {Reynoso}}, \bibinfo {author} {\bibfnamefont {G.}~\bibnamefont {Usaj}}, \
  and\ \bibinfo {author} {\bibfnamefont {C.~A.}\ \bibnamefont {Balseiro}},\
  }\href {\doibase 10.1103/PhysRevB.78.115312} {\bibfield  {journal} {\bibinfo
  {journal} {Phys. Rev. B}\ }\textbf {\bibinfo {volume} {78}},\ \bibinfo
  {pages} {115312} (\bibinfo {year} {2008}{\natexlab{a}})}\BibitemShut
  {NoStop}%
\bibitem [{\citenamefont {Korm\'anyos}(2010)}]{Kormanyos2010}%
  \BibitemOpen
  \bibfield  {author} {\bibinfo {author} {\bibfnamefont {A.}~\bibnamefont
  {Korm\'anyos}},\ }\href {\doibase 10.1103/PhysRevB.82.155316} {\bibfield
  {journal} {\bibinfo  {journal} {Phys. Rev. B}\ }\textbf {\bibinfo {volume}
  {82}},\ \bibinfo {pages} {155316} (\bibinfo {year} {2010})}\BibitemShut
  {NoStop}%
\bibitem [{\citenamefont {Bladwell}\ and\ \citenamefont
  {Sushkov}(2015)}]{Bladwell2015}%
  \BibitemOpen
  \bibfield  {author} {\bibinfo {author} {\bibfnamefont {S.}~\bibnamefont
  {Bladwell}}\ and\ \bibinfo {author} {\bibfnamefont {O.~P.}\ \bibnamefont
  {Sushkov}},\ }\href {\doibase 10.1103/PhysRevB.92.235416} {\bibfield
  {journal} {\bibinfo  {journal} {Phys. Rev. B}\ }\textbf {\bibinfo {volume}
  {92}},\ \bibinfo {pages} {235416} (\bibinfo {year} {2015})}\BibitemShut
  {NoStop}%
\bibitem [{\citenamefont {Yan}\ \emph {et~al.}(2018{\natexlab{a}})\citenamefont
  {Yan}, \citenamefont {Kumar}, \citenamefont {Pepper}, \citenamefont {Thomas},
  \citenamefont {See}, \citenamefont {Farrer}, \citenamefont {Ritchie},
  \citenamefont {Griffiths},\ and\ \citenamefont {Jones}}]{Yan2018}%
  \BibitemOpen
  \bibfield  {author} {\bibinfo {author} {\bibfnamefont {C.}~\bibnamefont
  {Yan}}, \bibinfo {author} {\bibfnamefont {S.}~\bibnamefont {Kumar}}, \bibinfo
  {author} {\bibfnamefont {M.}~\bibnamefont {Pepper}}, \bibinfo {author}
  {\bibfnamefont {K.}~\bibnamefont {Thomas}}, \bibinfo {author} {\bibfnamefont
  {P.}~\bibnamefont {See}}, \bibinfo {author} {\bibfnamefont {I.}~\bibnamefont
  {Farrer}}, \bibinfo {author} {\bibfnamefont {D.}~\bibnamefont {Ritchie}},
  \bibinfo {author} {\bibfnamefont {J.}~\bibnamefont {Griffiths}}, \ and\
  \bibinfo {author} {\bibfnamefont {G.}~\bibnamefont {Jones}},\ }\href
  {http://stacks.iop.org/1742-6596/964/i=1/a=012002} {\bibfield  {journal}
  {\bibinfo  {journal} {J. Phys. Conf. Ser.}\ }\textbf {\bibinfo {volume}
  {964}},\ \bibinfo {pages} {012002} (\bibinfo {year}
  {2018}{\natexlab{a}})}\BibitemShut {NoStop}%
\bibitem [{\citenamefont {Kohda}\ \emph {et~al.}(2012)\citenamefont {Kohda},
  \citenamefont {Nakamura}, \citenamefont {Nishihara}, \citenamefont
  {Kobayashi}, \citenamefont {Ono}, \citenamefont {Ohe}, \citenamefont
  {Tokura}, \citenamefont {Mineno},\ and\ \citenamefont {Nitta}}]{Kohda2012}%
  \BibitemOpen
  \bibfield  {author} {\bibinfo {author} {\bibfnamefont {M.}~\bibnamefont
  {Kohda}}, \bibinfo {author} {\bibfnamefont {S.}~\bibnamefont {Nakamura}},
  \bibinfo {author} {\bibfnamefont {Y.}~\bibnamefont {Nishihara}}, \bibinfo
  {author} {\bibfnamefont {K.}~\bibnamefont {Kobayashi}}, \bibinfo {author}
  {\bibfnamefont {T.}~\bibnamefont {Ono}}, \bibinfo {author} {\bibfnamefont
  {J.-i.}\ \bibnamefont {Ohe}}, \bibinfo {author} {\bibfnamefont
  {Y.}~\bibnamefont {Tokura}}, \bibinfo {author} {\bibfnamefont
  {T.}~\bibnamefont {Mineno}}, \ and\ \bibinfo {author} {\bibfnamefont
  {J.}~\bibnamefont {Nitta}},\ }\href {http://dx.doi.org/10.1038/ncomms2080}
  {\bibfield  {journal} {\bibinfo  {journal} {Nature Communications}\ }\textbf
  {\bibinfo {volume} {3}},\ \bibinfo {pages} {1082} (\bibinfo {year}
  {2012})}\BibitemShut {NoStop}%
\bibitem [{\citenamefont {Zeng}\ and\ \citenamefont {Liang}(2012)}]{Zeng2012}%
  \BibitemOpen
  \bibfield  {author} {\bibinfo {author} {\bibfnamefont {M.}~\bibnamefont
  {Zeng}}\ and\ \bibinfo {author} {\bibfnamefont {G.}~\bibnamefont {Liang}},\
  }\href {\doibase 10.1063/1.4757411} {\bibfield  {journal} {\bibinfo
  {journal} {Journal of Applied Physics}\ }\textbf {\bibinfo {volume} {112}},\
  \bibinfo {pages} {073707} (\bibinfo {year} {2012})}\BibitemShut {NoStop}%
\bibitem [{\citenamefont {Gupta}\ \emph {et~al.}(2014)\citenamefont {Gupta},
  \citenamefont {Lin}, \citenamefont {Bansil}, \citenamefont {Abdul~Jalil},
  \citenamefont {Huang}, \citenamefont {Tsai},\ and\ \citenamefont
  {Liang}}]{Gupta2014}%
  \BibitemOpen
  \bibfield  {author} {\bibinfo {author} {\bibfnamefont {G.}~\bibnamefont
  {Gupta}}, \bibinfo {author} {\bibfnamefont {H.}~\bibnamefont {Lin}}, \bibinfo
  {author} {\bibfnamefont {A.}~\bibnamefont {Bansil}}, \bibinfo {author}
  {\bibfnamefont {M.~B.}\ \bibnamefont {Abdul~Jalil}}, \bibinfo {author}
  {\bibfnamefont {C.-Y.}\ \bibnamefont {Huang}}, \bibinfo {author}
  {\bibfnamefont {W.-F.}\ \bibnamefont {Tsai}}, \ and\ \bibinfo {author}
  {\bibfnamefont {G.}~\bibnamefont {Liang}},\ }\href {\doibase
  10.1063/1.4863088} {\bibfield  {journal} {\bibinfo  {journal} {Applied
  Physics Letters}\ }\textbf {\bibinfo {volume} {104}},\ \bibinfo {pages}
  {032410} (\bibinfo {year} {2014})}\BibitemShut {NoStop}%
\bibitem [{\citenamefont {Meier}\ \emph {et~al.}(2007)\citenamefont {Meier},
  \citenamefont {Salis}, \citenamefont {Shorubalko}, \citenamefont {Gini},
  \citenamefont {Sch{\"o}n},\ and\ \citenamefont {Ensslin}}]{Meier2007}%
  \BibitemOpen
  \bibfield  {author} {\bibinfo {author} {\bibfnamefont {L.}~\bibnamefont
  {Meier}}, \bibinfo {author} {\bibfnamefont {G.}~\bibnamefont {Salis}},
  \bibinfo {author} {\bibfnamefont {I.}~\bibnamefont {Shorubalko}}, \bibinfo
  {author} {\bibfnamefont {E.}~\bibnamefont {Gini}}, \bibinfo {author}
  {\bibfnamefont {S.}~\bibnamefont {Sch{\"o}n}}, \ and\ \bibinfo {author}
  {\bibfnamefont {K.}~\bibnamefont {Ensslin}},\ }\href
  {http://dx.doi.org/10.1038/nphys675} {\bibfield  {journal} {\bibinfo
  {journal} {Nature Physics}\ }\textbf {\bibinfo {volume} {3}},\ \bibinfo
  {pages} {650} (\bibinfo {year} {2007})}\BibitemShut {NoStop}%
\bibitem [{\citenamefont {Reynoso}\ \emph
  {et~al.}(2008{\natexlab{b}})\citenamefont {Reynoso}, \citenamefont {Usaj},\
  and\ \citenamefont {Balseiro}}]{Reynoso2008}%
  \BibitemOpen
  \bibfield  {author} {\bibinfo {author} {\bibfnamefont {A.}~\bibnamefont
  {Reynoso}}, \bibinfo {author} {\bibfnamefont {G.}~\bibnamefont {Usaj}}, \
  and\ \bibinfo {author} {\bibfnamefont {C.}~\bibnamefont {Balseiro}},\ }in\
  \href {\doibase https://doi.org/10.1007/978-1-4020-8512-3_11} {\emph
  {\bibinfo {booktitle} {Quantum Magnetism}}},\ \bibinfo {editor} {edited by\
  \bibinfo {editor} {\bibfnamefont {B.}~\bibnamefont {Barbara}}, \bibinfo
  {editor} {\bibfnamefont {Y.}~\bibnamefont {Imry}}, \bibinfo {editor}
  {\bibfnamefont {G.}~\bibnamefont {Sawatzky}}, \ and\ \bibinfo {editor}
  {\bibfnamefont {P.}~\bibnamefont {Stamp}}}\ (\bibinfo  {publisher} {Springer,
  Dordrecht},\ \bibinfo {year} {2008})\ p.\ \bibinfo {pages} {151}\BibitemShut
  {NoStop}%
\bibitem [{\citenamefont {Watson}\ \emph {et~al.}(2003)\citenamefont {Watson},
  \citenamefont {Potok}, \citenamefont {Marcus},\ and\ \citenamefont
  {Umansky}}]{Watson2003}%
  \BibitemOpen
  \bibfield  {author} {\bibinfo {author} {\bibfnamefont {S.~K.}\ \bibnamefont
  {Watson}}, \bibinfo {author} {\bibfnamefont {R.~M.}\ \bibnamefont {Potok}},
  \bibinfo {author} {\bibfnamefont {C.~M.}\ \bibnamefont {Marcus}}, \ and\
  \bibinfo {author} {\bibfnamefont {V.}~\bibnamefont {Umansky}},\ }\href
  {\doibase 10.1103/PhysRevLett.91.258301} {\bibfield  {journal} {\bibinfo
  {journal} {Phys. Rev. Lett.}\ }\textbf {\bibinfo {volume} {91}},\ \bibinfo
  {pages} {258301} (\bibinfo {year} {2003})}\BibitemShut {NoStop}%
\bibitem [{\citenamefont {Li}\ \emph {et~al.}(2012)\citenamefont {Li},
  \citenamefont {Gilbertson}, \citenamefont {Litvinenko}, \citenamefont
  {Cohen},\ and\ \citenamefont {Clowes}}]{Li2012}%
  \BibitemOpen
  \bibfield  {author} {\bibinfo {author} {\bibfnamefont {J.}~\bibnamefont
  {Li}}, \bibinfo {author} {\bibfnamefont {A.~M.}\ \bibnamefont {Gilbertson}},
  \bibinfo {author} {\bibfnamefont {K.~L.}\ \bibnamefont {Litvinenko}},
  \bibinfo {author} {\bibfnamefont {L.~F.}\ \bibnamefont {Cohen}}, \ and\
  \bibinfo {author} {\bibfnamefont {S.~K.}\ \bibnamefont {Clowes}},\ }\href
  {\doibase 10.1103/PhysRevB.85.045431} {\bibfield  {journal} {\bibinfo
  {journal} {Phys. Rev. B}\ }\textbf {\bibinfo {volume} {85}},\ \bibinfo
  {pages} {045431} (\bibinfo {year} {2012})}\BibitemShut {NoStop}%
\bibitem [{\citenamefont {Yan}\ \emph {et~al.}(2018{\natexlab{b}})\citenamefont
  {Yan}, \citenamefont {Kumar}, \citenamefont {Thomas}, \citenamefont {See},
  \citenamefont {Farrer}, \citenamefont {Ritchie}, \citenamefont {Griffiths},
  \citenamefont {Jones},\ and\ \citenamefont {Pepper}}]{Yan2018a}%
  \BibitemOpen
  \bibfield  {author} {\bibinfo {author} {\bibfnamefont {C.}~\bibnamefont
  {Yan}}, \bibinfo {author} {\bibfnamefont {S.}~\bibnamefont {Kumar}}, \bibinfo
  {author} {\bibfnamefont {K.}~\bibnamefont {Thomas}}, \bibinfo {author}
  {\bibfnamefont {P.}~\bibnamefont {See}}, \bibinfo {author} {\bibfnamefont
  {I.}~\bibnamefont {Farrer}}, \bibinfo {author} {\bibfnamefont
  {D.}~\bibnamefont {Ritchie}}, \bibinfo {author} {\bibfnamefont
  {J.}~\bibnamefont {Griffiths}}, \bibinfo {author} {\bibfnamefont
  {G.}~\bibnamefont {Jones}}, \ and\ \bibinfo {author} {\bibfnamefont
  {M.}~\bibnamefont {Pepper}},\ }\href
  {http://stacks.iop.org/0953-8984/30/i=8/a=08LT01} {\bibfield  {journal}
  {\bibinfo  {journal} {J. Phys. Condens. Matter}\ }\textbf {\bibinfo {volume}
  {30}},\ \bibinfo {pages} {08LT01} (\bibinfo {year}
  {2018}{\natexlab{b}})}\BibitemShut {NoStop}%
\bibitem [{\citenamefont {Wharam}\ \emph {et~al.}(1988)\citenamefont {Wharam},
  \citenamefont {Thornton}, \citenamefont {Newbury}, \citenamefont {Pepper},
  \citenamefont {Ahmed}, \citenamefont {Frost}, \citenamefont {Hasko},
  \citenamefont {Peacock}, \citenamefont {Ritchie},\ and\ \citenamefont
  {Jones}}]{Wharam1988}%
  \BibitemOpen
  \bibfield  {author} {\bibinfo {author} {\bibfnamefont {D.~A.}\ \bibnamefont
  {Wharam}}, \bibinfo {author} {\bibfnamefont {T.~J.}\ \bibnamefont
  {Thornton}}, \bibinfo {author} {\bibfnamefont {R.}~\bibnamefont {Newbury}},
  \bibinfo {author} {\bibfnamefont {M.}~\bibnamefont {Pepper}}, \bibinfo
  {author} {\bibfnamefont {H.}~\bibnamefont {Ahmed}}, \bibinfo {author}
  {\bibfnamefont {J.~E.~F.}\ \bibnamefont {Frost}}, \bibinfo {author}
  {\bibfnamefont {D.~G.}\ \bibnamefont {Hasko}}, \bibinfo {author}
  {\bibfnamefont {D.~C.}\ \bibnamefont {Peacock}}, \bibinfo {author}
  {\bibfnamefont {D.~A.}\ \bibnamefont {Ritchie}}, \ and\ \bibinfo {author}
  {\bibfnamefont {G.~A.~C.}\ \bibnamefont {Jones}},\ }\href
  {http://stacks.iop.org/0022-3719/21/i=8/a=002} {\bibfield  {journal}
  {\bibinfo  {journal} {J. Phys. C}\ }\textbf {\bibinfo {volume} {21}},\
  \bibinfo {pages} {L209} (\bibinfo {year} {1988})}\BibitemShut {NoStop}%
\bibitem [{\citenamefont {van Wees}\ \emph {et~al.}(1988)\citenamefont {van
  Wees}, \citenamefont {van Houten}, \citenamefont {Beenakker}, \citenamefont
  {Williamson}, \citenamefont {Kouwenhoven}, \citenamefont {van~der Marel},\
  and\ \citenamefont {Foxon}}]{Wees1988}%
  \BibitemOpen
  \bibfield  {author} {\bibinfo {author} {\bibfnamefont {B.~J.}\ \bibnamefont
  {van Wees}}, \bibinfo {author} {\bibfnamefont {H.}~\bibnamefont {van
  Houten}}, \bibinfo {author} {\bibfnamefont {C.~W.~J.}\ \bibnamefont
  {Beenakker}}, \bibinfo {author} {\bibfnamefont {J.~G.}\ \bibnamefont
  {Williamson}}, \bibinfo {author} {\bibfnamefont {L.~P.}\ \bibnamefont
  {Kouwenhoven}}, \bibinfo {author} {\bibfnamefont {D.}~\bibnamefont {van~der
  Marel}}, \ and\ \bibinfo {author} {\bibfnamefont {C.~T.}\ \bibnamefont
  {Foxon}},\ }\href {\doibase 10.1103/PhysRevLett.60.848} {\bibfield  {journal}
  {\bibinfo  {journal} {Phys. Rev. Lett.}\ }\textbf {\bibinfo {volume} {60}},\
  \bibinfo {pages} {848} (\bibinfo {year} {1988})}\BibitemShut {NoStop}%
\bibitem [{\citenamefont {van Wees}\ \emph {et~al.}(1991)\citenamefont {van
  Wees}, \citenamefont {Kouwenhoven}, \citenamefont {Willems}, \citenamefont
  {Harmans}, \citenamefont {Mooij}, \citenamefont {van Houten}, \citenamefont
  {Beenakker}, \citenamefont {Williamson},\ and\ \citenamefont
  {Foxon}}]{Wees1991}%
  \BibitemOpen
  \bibfield  {author} {\bibinfo {author} {\bibfnamefont {B.~J.}\ \bibnamefont
  {van Wees}}, \bibinfo {author} {\bibfnamefont {L.~P.}\ \bibnamefont
  {Kouwenhoven}}, \bibinfo {author} {\bibfnamefont {E.~M.~M.}\ \bibnamefont
  {Willems}}, \bibinfo {author} {\bibfnamefont {C.~J. P.~M.}\ \bibnamefont
  {Harmans}}, \bibinfo {author} {\bibfnamefont {J.~E.}\ \bibnamefont {Mooij}},
  \bibinfo {author} {\bibfnamefont {H.}~\bibnamefont {van Houten}}, \bibinfo
  {author} {\bibfnamefont {C.~W.~J.}\ \bibnamefont {Beenakker}}, \bibinfo
  {author} {\bibfnamefont {J.~G.}\ \bibnamefont {Williamson}}, \ and\ \bibinfo
  {author} {\bibfnamefont {C.~T.}\ \bibnamefont {Foxon}},\ }\href {\doibase
  10.1103/PhysRevB.43.12431} {\bibfield  {journal} {\bibinfo  {journal} {Phys.
  Rev. B}\ }\textbf {\bibinfo {volume} {43}},\ \bibinfo {pages} {12431}
  (\bibinfo {year} {1991})}\BibitemShut {NoStop}%
\bibitem [{\citenamefont {Qu}\ \emph {et~al.}(2016)\citenamefont {Qu},
  \citenamefont {van Veen}, \citenamefont {de~Vries}, \citenamefont {Beukman},
  \citenamefont {Wimmer}, \citenamefont {Yi}, \citenamefont {Kiselev},
  \citenamefont {Nguyen}, \citenamefont {Sokolich}, \citenamefont {Manfra},
  \citenamefont {Nichele}, \citenamefont {Marcus},\ and\ \citenamefont
  {Kouwenhoven}}]{Qu2016}%
  \BibitemOpen
  \bibfield  {author} {\bibinfo {author} {\bibfnamefont {F.}~\bibnamefont
  {Qu}}, \bibinfo {author} {\bibfnamefont {J.}~\bibnamefont {van Veen}},
  \bibinfo {author} {\bibfnamefont {F.~K.}\ \bibnamefont {de~Vries}}, \bibinfo
  {author} {\bibfnamefont {A.~J.~A.}\ \bibnamefont {Beukman}}, \bibinfo
  {author} {\bibfnamefont {M.}~\bibnamefont {Wimmer}}, \bibinfo {author}
  {\bibfnamefont {W.}~\bibnamefont {Yi}}, \bibinfo {author} {\bibfnamefont
  {A.~A.}\ \bibnamefont {Kiselev}}, \bibinfo {author} {\bibfnamefont {B.-M.}\
  \bibnamefont {Nguyen}}, \bibinfo {author} {\bibfnamefont {M.}~\bibnamefont
  {Sokolich}}, \bibinfo {author} {\bibfnamefont {M.~J.}\ \bibnamefont
  {Manfra}}, \bibinfo {author} {\bibfnamefont {F.}~\bibnamefont {Nichele}},
  \bibinfo {author} {\bibfnamefont {C.~M.}\ \bibnamefont {Marcus}}, \ and\
  \bibinfo {author} {\bibfnamefont {L.~P.}\ \bibnamefont {Kouwenhoven}},\
  }\href {\doibase 10.1021/acs.nanolett.6b03297} {\bibfield  {journal}
  {\bibinfo  {journal} {Nano Lett.}\ }\textbf {\bibinfo {volume} {16}},\
  \bibinfo {pages} {7509} (\bibinfo {year} {2016})}\BibitemShut {NoStop}%
\bibitem [{\citenamefont {LeRoy}\ \emph {et~al.}(2005)\citenamefont {LeRoy},
  \citenamefont {Bleszynski}, \citenamefont {Aidala}, \citenamefont
  {Westervelt}, \citenamefont {Kalben}, \citenamefont {Heller}, \citenamefont
  {Shaw}, \citenamefont {Maranowski},\ and\ \citenamefont
  {Gossard}}]{LeRoy2005}%
  \BibitemOpen
  \bibfield  {author} {\bibinfo {author} {\bibfnamefont {B.~J.}\ \bibnamefont
  {LeRoy}}, \bibinfo {author} {\bibfnamefont {A.~C.}\ \bibnamefont
  {Bleszynski}}, \bibinfo {author} {\bibfnamefont {K.~E.}\ \bibnamefont
  {Aidala}}, \bibinfo {author} {\bibfnamefont {R.~M.}\ \bibnamefont
  {Westervelt}}, \bibinfo {author} {\bibfnamefont {A.}~\bibnamefont {Kalben}},
  \bibinfo {author} {\bibfnamefont {E.~J.}\ \bibnamefont {Heller}}, \bibinfo
  {author} {\bibfnamefont {S.~E.~J.}\ \bibnamefont {Shaw}}, \bibinfo {author}
  {\bibfnamefont {K.~D.}\ \bibnamefont {Maranowski}}, \ and\ \bibinfo {author}
  {\bibfnamefont {A.~C.}\ \bibnamefont {Gossard}},\ }\href {\doibase
  10.1103/PhysRevLett.94.126801} {\bibfield  {journal} {\bibinfo  {journal}
  {Phys. Rev. Lett.}\ }\textbf {\bibinfo {volume} {94}},\ \bibinfo {pages}
  {126801} (\bibinfo {year} {2005})}\BibitemShut {NoStop}%
\bibitem [{\citenamefont {Jura}\ \emph {et~al.}(2009)\citenamefont {Jura},
  \citenamefont {Topinka}, \citenamefont {Grobis}, \citenamefont {Pfeiffer},
  \citenamefont {West},\ and\ \citenamefont {Goldhaber-Gordon}}]{Jura2009}%
  \BibitemOpen
  \bibfield  {author} {\bibinfo {author} {\bibfnamefont {M.~P.}\ \bibnamefont
  {Jura}}, \bibinfo {author} {\bibfnamefont {M.~A.}\ \bibnamefont {Topinka}},
  \bibinfo {author} {\bibfnamefont {M.}~\bibnamefont {Grobis}}, \bibinfo
  {author} {\bibfnamefont {L.~N.}\ \bibnamefont {Pfeiffer}}, \bibinfo {author}
  {\bibfnamefont {K.~W.}\ \bibnamefont {West}}, \ and\ \bibinfo {author}
  {\bibfnamefont {D.}~\bibnamefont {Goldhaber-Gordon}},\ }\href {\doibase
  10.1103/PhysRevB.80.041303} {\bibfield  {journal} {\bibinfo  {journal} {Phys.
  Rev. B}\ }\textbf {\bibinfo {volume} {80}},\ \bibinfo {pages} {041303}
  (\bibinfo {year} {2009})}\BibitemShut {NoStop}%
\bibitem [{\citenamefont {Paradiso}\ \emph {et~al.}(2010)\citenamefont
  {Paradiso}, \citenamefont {Heun}, \citenamefont {Roddaro}, \citenamefont
  {Pfeiffer}, \citenamefont {West}, \citenamefont {Sorba}, \citenamefont
  {Biasiol},\ and\ \citenamefont {Beltram}}]{Paradiso2010}%
  \BibitemOpen
  \bibfield  {author} {\bibinfo {author} {\bibfnamefont {N.}~\bibnamefont
  {Paradiso}}, \bibinfo {author} {\bibfnamefont {S.}~\bibnamefont {Heun}},
  \bibinfo {author} {\bibfnamefont {S.}~\bibnamefont {Roddaro}}, \bibinfo
  {author} {\bibfnamefont {L.}~\bibnamefont {Pfeiffer}}, \bibinfo {author}
  {\bibfnamefont {K.}~\bibnamefont {West}}, \bibinfo {author} {\bibfnamefont
  {L.}~\bibnamefont {Sorba}}, \bibinfo {author} {\bibfnamefont
  {G.}~\bibnamefont {Biasiol}}, \ and\ \bibinfo {author} {\bibfnamefont
  {F.}~\bibnamefont {Beltram}},\ }\href {\doibase
  https://doi.org/10.1016/j.physe.2009.11.146} {\bibfield  {journal} {\bibinfo
  {journal} {Physica E Low Dimens. Syst. Nanostruct.}\ }\textbf {\bibinfo
  {volume} {42}},\ \bibinfo {pages} {1038} (\bibinfo {year}
  {2010})}\BibitemShut {NoStop}%
\bibitem [{\citenamefont {Brun}\ \emph {et~al.}(2014)\citenamefont {Brun},
  \citenamefont {Martins}, \citenamefont {Faniel}, \citenamefont {Hackens},
  \citenamefont {Bachelier}, \citenamefont {Cavanna}, \citenamefont {Ulysse},
  \citenamefont {Ouerghi}, \citenamefont {Gennser}, \citenamefont {Mailly},
  \citenamefont {Huant}, \citenamefont {Bayot}, \citenamefont {Sanquer},\ and\
  \citenamefont {Sellier}}]{Brun2014}%
  \BibitemOpen
  \bibfield  {author} {\bibinfo {author} {\bibfnamefont {B.}~\bibnamefont
  {Brun}}, \bibinfo {author} {\bibfnamefont {F.}~\bibnamefont {Martins}},
  \bibinfo {author} {\bibfnamefont {S.}~\bibnamefont {Faniel}}, \bibinfo
  {author} {\bibfnamefont {B.}~\bibnamefont {Hackens}}, \bibinfo {author}
  {\bibfnamefont {G.}~\bibnamefont {Bachelier}}, \bibinfo {author}
  {\bibfnamefont {A.}~\bibnamefont {Cavanna}}, \bibinfo {author} {\bibfnamefont
  {C.}~\bibnamefont {Ulysse}}, \bibinfo {author} {\bibfnamefont
  {A.}~\bibnamefont {Ouerghi}}, \bibinfo {author} {\bibfnamefont
  {U.}~\bibnamefont {Gennser}}, \bibinfo {author} {\bibfnamefont
  {D.}~\bibnamefont {Mailly}}, \bibinfo {author} {\bibfnamefont
  {S.}~\bibnamefont {Huant}}, \bibinfo {author} {\bibfnamefont
  {V.}~\bibnamefont {Bayot}}, \bibinfo {author} {\bibfnamefont
  {M.}~\bibnamefont {Sanquer}}, \ and\ \bibinfo {author} {\bibfnamefont
  {H.}~\bibnamefont {Sellier}},\ }\href {http://dx.doi.org/10.1038/ncomms5290}
  {\bibfield  {journal} {\bibinfo  {journal} {Nat. Commun.}\ }\textbf {\bibinfo
  {volume} {5}},\ \bibinfo {pages} {4290} (\bibinfo {year} {2014})}\BibitemShut
  {NoStop}%
\bibitem [{\citenamefont {Crook}\ \emph {et~al.}(2003)\citenamefont {Crook},
  \citenamefont {Smith}, \citenamefont {Graham}, \citenamefont {Farrer},
  \citenamefont {Beere},\ and\ \citenamefont {Ritchie}}]{Crook2003}%
  \BibitemOpen
  \bibfield  {author} {\bibinfo {author} {\bibfnamefont {R.}~\bibnamefont
  {Crook}}, \bibinfo {author} {\bibfnamefont {C.~G.}\ \bibnamefont {Smith}},
  \bibinfo {author} {\bibfnamefont {A.~C.}\ \bibnamefont {Graham}}, \bibinfo
  {author} {\bibfnamefont {I.}~\bibnamefont {Farrer}}, \bibinfo {author}
  {\bibfnamefont {H.~E.}\ \bibnamefont {Beere}}, \ and\ \bibinfo {author}
  {\bibfnamefont {D.~A.}\ \bibnamefont {Ritchie}},\ }\href {\doibase
  10.1103/PhysRevLett.91.246803} {\bibfield  {journal} {\bibinfo  {journal}
  {Phys. Rev. Lett.}\ }\textbf {\bibinfo {volume} {91}},\ \bibinfo {pages}
  {246803} (\bibinfo {year} {2003})}\BibitemShut {NoStop}%
\bibitem [{\citenamefont {Burke}\ \emph {et~al.}(2010)\citenamefont {Burke},
  \citenamefont {Akis}, \citenamefont {Day}, \citenamefont {Speyer},
  \citenamefont {Ferry},\ and\ \citenamefont {Bennett}}]{Burke2010}%
  \BibitemOpen
  \bibfield  {author} {\bibinfo {author} {\bibfnamefont {A.~M.}\ \bibnamefont
  {Burke}}, \bibinfo {author} {\bibfnamefont {R.}~\bibnamefont {Akis}},
  \bibinfo {author} {\bibfnamefont {T.~E.}\ \bibnamefont {Day}}, \bibinfo
  {author} {\bibfnamefont {G.}~\bibnamefont {Speyer}}, \bibinfo {author}
  {\bibfnamefont {D.~K.}\ \bibnamefont {Ferry}}, \ and\ \bibinfo {author}
  {\bibfnamefont {B.~R.}\ \bibnamefont {Bennett}},\ }\href {\doibase
  10.1103/PhysRevLett.104.176801} {\bibfield  {journal} {\bibinfo  {journal}
  {Phys. Rev. Lett.}\ }\textbf {\bibinfo {volume} {104}},\ \bibinfo {pages}
  {176801} (\bibinfo {year} {2010})}\BibitemShut {NoStop}%
\bibitem [{\citenamefont {Crook}\ \emph {et~al.}(2000)\citenamefont {Crook},
  \citenamefont {Smith}, \citenamefont {Simmons},\ and\ \citenamefont
  {Ritchie}}]{Crook2010}%
  \BibitemOpen
  \bibfield  {author} {\bibinfo {author} {\bibfnamefont {R.}~\bibnamefont
  {Crook}}, \bibinfo {author} {\bibfnamefont {C.~G.}\ \bibnamefont {Smith}},
  \bibinfo {author} {\bibfnamefont {M.~Y.}\ \bibnamefont {Simmons}}, \ and\
  \bibinfo {author} {\bibfnamefont {D.~A.}\ \bibnamefont {Ritchie}},\ }\href
  {\doibase 10.1103/PhysRevB.62.5174} {\bibfield  {journal} {\bibinfo
  {journal} {Phys. Rev. B}\ }\textbf {\bibinfo {volume} {62}},\ \bibinfo
  {pages} {5174} (\bibinfo {year} {2000})}\BibitemShut {NoStop}%
\bibitem [{\citenamefont {Morikawa}\ \emph {et~al.}(2015)\citenamefont
  {Morikawa}, \citenamefont {Dou}, \citenamefont {Wang}, \citenamefont {Smith},
  \citenamefont {Watanabe}, \citenamefont {Taniguchi}, \citenamefont
  {Masubuchi}, \citenamefont {Machida},\ and\ \citenamefont
  {Connolly}}]{Morikawa2015}%
  \BibitemOpen
  \bibfield  {author} {\bibinfo {author} {\bibfnamefont {S.}~\bibnamefont
  {Morikawa}}, \bibinfo {author} {\bibfnamefont {Z.}~\bibnamefont {Dou}},
  \bibinfo {author} {\bibfnamefont {S.-W.}\ \bibnamefont {Wang}}, \bibinfo
  {author} {\bibfnamefont {C.~G.}\ \bibnamefont {Smith}}, \bibinfo {author}
  {\bibfnamefont {K.}~\bibnamefont {Watanabe}}, \bibinfo {author}
  {\bibfnamefont {T.}~\bibnamefont {Taniguchi}}, \bibinfo {author}
  {\bibfnamefont {S.}~\bibnamefont {Masubuchi}}, \bibinfo {author}
  {\bibfnamefont {T.}~\bibnamefont {Machida}}, \ and\ \bibinfo {author}
  {\bibfnamefont {M.~R.}\ \bibnamefont {Connolly}},\ }\href {\doibase
  10.1063/1.4937473} {\bibfield  {journal} {\bibinfo  {journal} {Appl. Phys.
  Lett.}\ }\textbf {\bibinfo {volume} {107}},\ \bibinfo {pages} {243102}
  (\bibinfo {year} {2015})}\BibitemShut {NoStop}%
\bibitem [{\citenamefont {Bhandari}\ \emph {et~al.}(2016)\citenamefont
  {Bhandari}, \citenamefont {Lee}, \citenamefont {Klales}, \citenamefont
  {Watanabe}, \citenamefont {Taniguchi}, \citenamefont {Heller}, \citenamefont
  {Kim},\ and\ \citenamefont {Westervelt}}]{Bhandari2016}%
  \BibitemOpen
  \bibfield  {author} {\bibinfo {author} {\bibfnamefont {S.}~\bibnamefont
  {Bhandari}}, \bibinfo {author} {\bibfnamefont {G.-H.}\ \bibnamefont {Lee}},
  \bibinfo {author} {\bibfnamefont {A.}~\bibnamefont {Klales}}, \bibinfo
  {author} {\bibfnamefont {K.}~\bibnamefont {Watanabe}}, \bibinfo {author}
  {\bibfnamefont {T.}~\bibnamefont {Taniguchi}}, \bibinfo {author}
  {\bibfnamefont {E.}~\bibnamefont {Heller}}, \bibinfo {author} {\bibfnamefont
  {P.}~\bibnamefont {Kim}}, \ and\ \bibinfo {author} {\bibfnamefont {R.~M.}\
  \bibnamefont {Westervelt}},\ }\href {\doibase 10.1021/acs.nanolett.5b04609}
  {\bibfield  {journal} {\bibinfo  {journal} {Nano Lett.}\ }\textbf {\bibinfo
  {volume} {16}},\ \bibinfo {pages} {1690} (\bibinfo {year}
  {2016})}\BibitemShut {NoStop}%
\bibitem [{\citenamefont {Davies}\ \emph {et~al.}(1995)\citenamefont {Davies},
  \citenamefont {Larkin},\ and\ \citenamefont {Sukhorukov}}]{Davies1995}%
  \BibitemOpen
  \bibfield  {author} {\bibinfo {author} {\bibfnamefont {J.~H.}\ \bibnamefont
  {Davies}}, \bibinfo {author} {\bibfnamefont {I.~A.}\ \bibnamefont {Larkin}},
  \ and\ \bibinfo {author} {\bibfnamefont {E.~V.}\ \bibnamefont {Sukhorukov}},\
  }\href {\doibase 10.1063/1.359446} {\bibfield  {journal} {\bibinfo  {journal}
  {J. Appl. Phys.}\ }\textbf {\bibinfo {volume} {77}},\ \bibinfo {pages} {4504}
  (\bibinfo {year} {1995})}\BibitemShut {NoStop}%
\bibitem [{\citenamefont {Gilbertson}\ \emph {et~al.}(2008)\citenamefont
  {Gilbertson}, \citenamefont {Fearn}, \citenamefont {Jefferson}, \citenamefont
  {Murdin}, \citenamefont {Buckle},\ and\ \citenamefont
  {Cohen}}]{Gilbertson2008}%
  \BibitemOpen
  \bibfield  {author} {\bibinfo {author} {\bibfnamefont {A.~M.}\ \bibnamefont
  {Gilbertson}}, \bibinfo {author} {\bibfnamefont {M.}~\bibnamefont {Fearn}},
  \bibinfo {author} {\bibfnamefont {J.~H.}\ \bibnamefont {Jefferson}}, \bibinfo
  {author} {\bibfnamefont {B.~N.}\ \bibnamefont {Murdin}}, \bibinfo {author}
  {\bibfnamefont {P.~D.}\ \bibnamefont {Buckle}}, \ and\ \bibinfo {author}
  {\bibfnamefont {L.~F.}\ \bibnamefont {Cohen}},\ }\href {\doibase
  10.1103/PhysRevB.77.165335} {\bibfield  {journal} {\bibinfo  {journal} {Phys.
  Rev. B}\ }\textbf {\bibinfo {volume} {77}},\ \bibinfo {pages} {165335}
  (\bibinfo {year} {2008})}\BibitemShut {NoStop}%
\bibitem [{\citenamefont {Kolasi\'{n}ski}\ \emph {et~al.}(2016)\citenamefont
  {Kolasi\'{n}ski}, \citenamefont {Szafran}, \citenamefont {Brun},\ and\
  \citenamefont {Sellier}}]{Kolacha}%
  \BibitemOpen
  \bibfield  {author} {\bibinfo {author} {\bibfnamefont {K.}~\bibnamefont
  {Kolasi\'{n}ski}}, \bibinfo {author} {\bibfnamefont {B.}~\bibnamefont
  {Szafran}}, \bibinfo {author} {\bibfnamefont {B.}~\bibnamefont {Brun}}, \
  and\ \bibinfo {author} {\bibfnamefont {H.}~\bibnamefont {Sellier}},\ }\href
  {\doibase 10.1103/PhysRevB.94.075301} {\bibfield  {journal} {\bibinfo
  {journal} {Phys. Rev. B}\ }\textbf {\bibinfo {volume} {94}},\ \bibinfo
  {pages} {075301} (\bibinfo {year} {2016})}\BibitemShut {NoStop}%
\bibitem [{\citenamefont {Potok}\ \emph {et~al.}(2002)\citenamefont {Potok},
  \citenamefont {Folk}, \citenamefont {Marcus},\ and\ \citenamefont
  {Umansky}}]{Potok2002}%
  \BibitemOpen
  \bibfield  {author} {\bibinfo {author} {\bibfnamefont {R.~M.}\ \bibnamefont
  {Potok}}, \bibinfo {author} {\bibfnamefont {J.~A.}\ \bibnamefont {Folk}},
  \bibinfo {author} {\bibfnamefont {C.~M.}\ \bibnamefont {Marcus}}, \ and\
  \bibinfo {author} {\bibfnamefont {V.}~\bibnamefont {Umansky}},\ }\href
  {\doibase 10.1103/PhysRevLett.89.266602} {\bibfield  {journal} {\bibinfo
  {journal} {Phys. Rev. Lett.}\ }\textbf {\bibinfo {volume} {89}},\ \bibinfo
  {pages} {266602} (\bibinfo {year} {2002})}\BibitemShut {NoStop}%
\bibitem [{\citenamefont {Plaut}\ \emph {et~al.}(1988)\citenamefont {Plaut},
  \citenamefont {Singleton}, \citenamefont {Nicholas}, \citenamefont {Harley},
  \citenamefont {Andrews},\ and\ \citenamefont {Foxon}}]{Plaut1988}%
  \BibitemOpen
  \bibfield  {author} {\bibinfo {author} {\bibfnamefont {A.~S.}\ \bibnamefont
  {Plaut}}, \bibinfo {author} {\bibfnamefont {J.}~\bibnamefont {Singleton}},
  \bibinfo {author} {\bibfnamefont {R.~J.}\ \bibnamefont {Nicholas}}, \bibinfo
  {author} {\bibfnamefont {R.~T.}\ \bibnamefont {Harley}}, \bibinfo {author}
  {\bibfnamefont {S.~R.}\ \bibnamefont {Andrews}}, \ and\ \bibinfo {author}
  {\bibfnamefont {C.~T.~B.}\ \bibnamefont {Foxon}},\ }\href {\doibase
  10.1103/PhysRevB.38.1323} {\bibfield  {journal} {\bibinfo  {journal} {Phys.
  Rev. B}\ }\textbf {\bibinfo {volume} {38}},\ \bibinfo {pages} {1323}
  (\bibinfo {year} {1988})}\BibitemShut {NoStop}%
\bibitem [{\citenamefont {Arora}\ \emph {et~al.}(2013)\citenamefont {Arora},
  \citenamefont {Mandal}, \citenamefont {Chakrabarti},\ and\ \citenamefont
  {Ghosh}}]{Arora2013}%
  \BibitemOpen
  \bibfield  {author} {\bibinfo {author} {\bibfnamefont {A.}~\bibnamefont
  {Arora}}, \bibinfo {author} {\bibfnamefont {A.}~\bibnamefont {Mandal}},
  \bibinfo {author} {\bibfnamefont {S.}~\bibnamefont {Chakrabarti}}, \ and\
  \bibinfo {author} {\bibfnamefont {S.}~\bibnamefont {Ghosh}},\ }\href
  {\doibase 10.1063/1.4808302} {\bibfield  {journal} {\bibinfo  {journal} {J.
  Appl. Phys.}\ }\textbf {\bibinfo {volume} {113}},\ \bibinfo {pages} {213505}
  (\bibinfo {year} {2013})}\BibitemShut {NoStop}%
\end{thebibliography}%

\end{document}